\documentclass[12pt]{article}

\usepackage{scicite}

\usepackage{amsmath,amssymb}
\usepackage{graphicx}
\usepackage{aecompl}
\usepackage{epstopdf}
\graphicspath{{.}{images/}}
\usepackage{dblfloatfix}
\usepackage{array}
\usepackage{multirow}
\usepackage{afterpage}

\usepackage{times}

\def\apj{ApJ}%
\def\apjl{ApJ}%
\def\apjs{ApJS}%
\def\aap{A\&A}%
\def\mnras{MNRAS}%
\def\pasp{PASP}%
\def\nat{Nature}%
\def\jqsrt{J.~Quant.~Spec.~Radiat.~Transf.}%
\newenvironment{sciabstract}{%
\begin{quote} \bf}
{\end{quote}}

\newcounter{lastnote}

\title{Constraining Exoplanet Mass from Transmission Spectroscopy$^\star$}

\author
{Julien de Wit$^{1\ast}$ and Sara Seager$^{1,2}$\\
\\
\normalsize{$^\star$This is the author's version of the work. It is posted here by permission of the
AAAS}\\
\normalsize{for personal use, not for redistribution. The definitive version was published in}\\
\normalsize{\textit{Science} (Vol. \textbf{342},  pp. 1473, 20 December 2013), DOI: \href{http://www.sciencemag.org/content/342/6165/1473}{10.1126/science.1245450}.} \\
\\
\normalsize{$^{1}$Department of Earth, Atmospheric and Planetary Sciences, Massachusetts Institute of Technology,}\\
\normalsize{77 Massachusetts Avenue, Cambridge, MA 02139, USA.}\\
\normalsize{$^{2}$Department of Physics, Massachusetts Institute of Technology,}\\
\normalsize{77 Massachusetts Avenue, Cambridge, MA 02139, USA.}\\
\normalsize{$^\ast$To whom correspondence should be addressed; E-mail:  jdewit@mit.edu.}
}

\date{}

\begin{document} 


\baselineskip24pt


\maketitle


\begin{sciabstract}

Determination of an exoplanet's mass is a key to understanding its basic properties, including its potential for supporting life. To date, mass constraints for exoplanets are predominantly based on radial velocity (RV) measurements, which are not suited for planets with low masses, large semi-major axes, or those orbiting faint or active stars. Here, we present a method to extract an exoplanet's mass solely from its transmission spectrum. We find good agreement between the mass retrieved for the hot Jupiter HD\,189733b from transmission spectroscopy with that from RV measurements. Our method will be able to retrieve the masses of Earth-sized and super-Earth planets using data from future space telescopes that were initially designed for atmospheric characterization.
\end{sciabstract}


\section{Introduction}
\label{sec:intro}

With over 900 confirmed exoplanets \cite{Schneider2011} and over 2300 planetary candidates known \cite{Batalha2013}, research priorities are moving from planet detection to planet characterization. In this context, a planet's mass is a fundamental parameter because it is connected to a planet's internal and atmospheric structure and it affects basic planetary processes such as the cooling of a planet, its plate tectonics \cite{Stamenkovic2012}, magnetic field generation, outgassing, and atmospheric escape. Measurement of a planetary mass can in many cases reveal the planet bulk composition, allowing to determine whether the planet is a gas giant or is rocky and suitable for life, as we know it. 

Planetary mass is traditionally constrained with the radial velocity (RV) technique using single-purpose dedicated instruments. The RV technique measures the Doppler shift of the stellar spectrum to derive the planet-to-star (minimum) mass ratio as the star orbits the planet-star common center of mass. Although the radial velocity technique has a pioneering history of success laying the foundation of the field of exoplanet detection, it is mainly effective for massive planets around relatively bright and quiet stars. Most transiting planets have host stars that are too faint for precise radial velocity measurements. For sufficiently bright host stars, stellar perturbations may be larger than the planet's signal, preventing a determination of the planet mass with radial velocity measurements even for hot Jupiters \cite{Collier2010}. In the long term, the limitation due to the faintness of targets will be reduced with technological improvements. However, host star perturbations may be a fundamental limit that cannot be overcome, meaning that the masses of small planets orbiting quiet stars would remain out of reach.
Current alternative mass measurements to RV are based on modulations of planetary-system light curves \cite{Mislis2012} or transit-timing variations \cite{Fabrycky2010}. The former works for massive planets on short period orbits and involves detection of both beaming and ellipsoidal modulations \cite{Faigler2011}. The latter relies on gravitational perturbations of a companion on the transiting planet's orbit. This method is most successful for companions that are themselves transiting and in orbital resonance with the planet of interest \cite{Agol2005,Holman2005}. For unseen companions the mass of the transiting planet is not constrained, but an upper limit on the mass of the unseen companion can be obtained to within 15 to 50\% \cite{Steffen2013}.

Transiting exoplanets are of special interest because the size of a transiting exoplanet can be derived from its transit light curve and combined with its mass, if known, to yield the planet's density, constraining its internal structure and potential habitability. Furthermore, the atmospheric properties of a transiting exoplanet can be retrieved from the host-star light passing through its atmosphere when it transits, but the quality of atmospheric retrieval is reduced if the planet's mass is inadequately constrained \cite{Barstow2013}. 

Here, we introduce \textit{MassSpec}, a method for constraining the mass of transiting exoplanets based solely on transit observations. \textit{MassSpec} extracts a planet's mass through its influence on the atmospheric scale height. It simultaneously and self-consistently constrains the mass and the atmosphere of an exoplanet, provides independent mass constraints for transiting planets accessible to RV, and allows us to determine the masses for transiting planets for which the radial velocity method fails. 

\section{\textit{MassSpec}: Concept and Feasibility}
\label{sec:spectrum2mass}

	The mass of a planet affects its transmission spectrum through the pressure profile of its atmosphere (i.e., $p(z)$ where $z$ is the altitude), and hence its atmospheric absorption profile. For an ideal gas atmosphere in hydrostatic equilibrium, the pressure varies with the altitude as $\text{d}\ln(p) = -\frac{1}{H} \text{d} z$, where $H$ is the atmospheric scale height defined as 		
\begin{eqnarray}
H & = & \frac{kT}{\mu g},
\label{sh1}
\end{eqnarray}
where $k$ is Boltzmann's constant and $T$, $\mu$ and $g$ are the local (i.e., altitude dependent) temperature, mean molecular mass, and gravity. By expressing the local gravity in terms of the planet's sizes (mass, $M_p$, and radius, $R_p$), Eq.\,\ref{sh1} can be rewritten as
\begin{equation}
M_p=\frac{kTR_p^2}{\mu GH}.
\label{M2}
\end{equation} 
Thus, our method conceptually requires constraining the radius of the target as well as its atmospheric temperature, mean molecular mass, and scale height. 

\begin{figure}[!ht]
  \begin{center}
    \vspace{-1cm}\includegraphics[width=6cm,height=!]{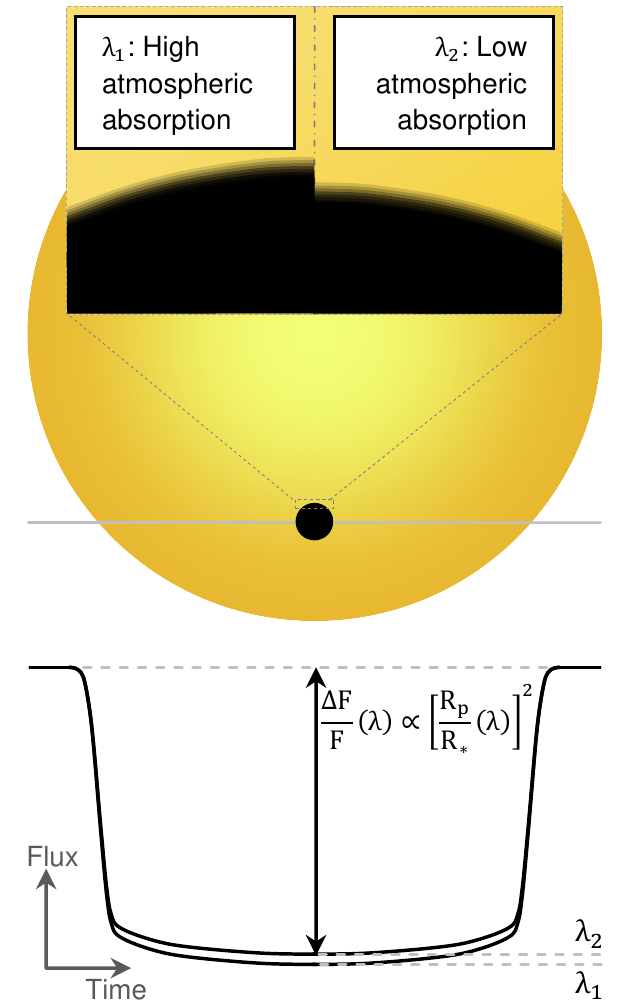}
  \end{center}
  \vspace{-0.6cm}
  \caption{Transit-depth variations, $ \frac{\Delta F}{F} (\lambda)$, induced by the wavelength-dependent opacity of a transiting planet atmosphere. The stellar disk and the planet are not resolved; the flux variation of a point source is observed.}
  \vspace{-1cm}
  \label{fig:transmission_spectrum}
\end{figure}

 A planet transmission spectrum can be seen as a wavelength-dependent drop in the apparent brightness of the host star when the planet transits (Fig.\ref{fig:transmission_spectrum}). At a wavelength with high atmospheric absorption, $\lambda_1$, the planet appears larger than at a wavelength with lower atmospheric absorption, $\lambda_2$---because of the additional flux drop due to the opaque atmospheric annulus. In particular, a relative flux-drop, $ \frac{\Delta F}{F} (\lambda)$, is associated with an apparent planet radius, $R_p(\lambda) = \sqrt{\frac{\Delta F}{F} (\lambda)} \times R_{\star}$. Transmission spectroscopy mainly probes low-pressure levels; therefore, the mass encompassed in the sphere of radius $R_p(\lambda)$ (Eq.\,\ref{M2}) is a good proxy for the planetary mass. 

\begin{figure}[!ht]
  \begin{center}
    \vspace{-0.5cm}\includegraphics[trim = 00mm 170mm 00mm 35mm,clip,width=17cm,height=!]{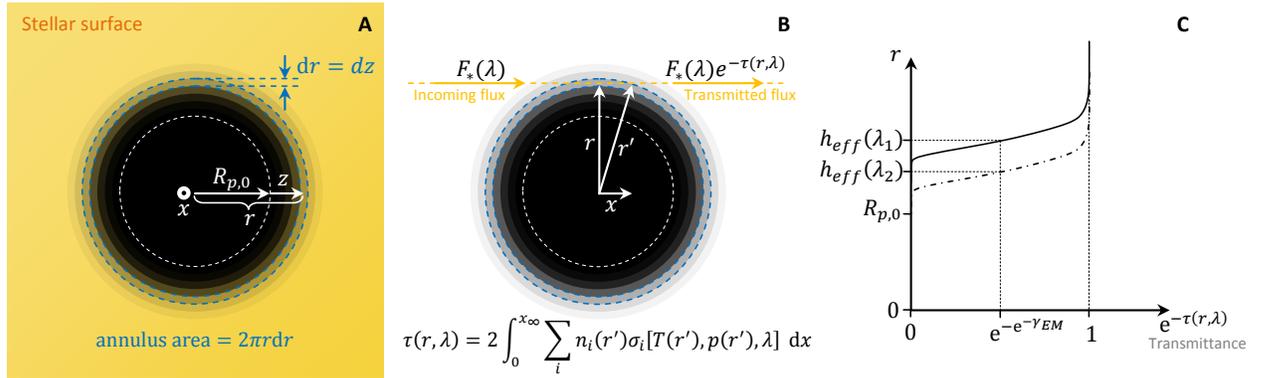}
  \end{center}
  \vspace{-0.6cm}
  \caption{Basics of a planet's transmission spectrum (planetary atmosphere scaled up to enhance visibility). (\textbf{A}) In-transit geometry as viewed by an observer presenting the areas of the atmospheric annulii affecting the transmission spectrum. (\textbf{B}) Side-view showing the flux transmitted through an atmospheric annulus of radius r. (\textbf{C}) Transmittance as a function of the radius at wavelengths with high and low atmospheric absorption---$\lambda_1$ (solid lines) and $\lambda_2$ (dash-dotted lines), respectively. Due to higher atmospheric absorption at $\lambda_1$, the planet will appear larger than it does at $\lambda_2$, because of the more-extended opaque atmospheric annulus $[h_{eff}(\lambda_1)) > h_{eff}(\lambda_2)]$ that translates into an additional flux drop \cite{science1note5}.}
  \vspace{-1cm}
  \label{fig:transmission_spectrum_basics}
\end{figure}

 The apparent radius of a planet relates directly to its atmospheric properties due to their effect on its opacity,
	\begin{eqnarray}
\pi R_p^2(\lambda) = \pi \left[R_{p,0} + h_{eff}(\lambda)\right]^2 & = & \int_{0}^{\infty} 2\pi r (1-e^{-\tau(r,\lambda)}) \text{ d}r,
\label{eq:transmission_spectrum_h}
\end{eqnarray}
where $R_{p,0}$, $h_{eff}(\lambda)$, and $e^{-\tau(r,\lambda)}$ are respectively a planetary radius of reference---i.e., any radial distance at which the body is optically thick in limb-looking over all the spectral band of interest, the effective atmosphere height, and the planet's transmittance at radius $r$ (Fig.\,\ref{fig:transmission_spectrum_basics}). $\tau(r,\lambda)$ is the slant-path optical depth defined as
\begin{eqnarray}
\tau(r,\lambda) & = & 2\int_{0}^{x_{\infty}} \sum_in_{i}(r') \times \sigma_{i}(T(r'),p(r'),\lambda) \text{ d}x,
 \label{eq:optical_depth}
\end{eqnarray} 
where $r' = \sqrt{r^2+x^2}$ and $n_{i}(r')$ and $\sigma_{i}(T(r'),p(r'),\lambda)$ are the number density and the extinction cross section of the $i^{th}$ atmospheric component at the radial distance $r'$ \cite{Seager2010}. In other words, a planet's atmospheric properties $[n_{i}(z),T(z),\mbox{ and }p(z)]$ are embedded in its transmission spectrum through $\tau(r,\lambda)$ (Eqs.\,\ref{eq:transmission_spectrum_h} and\,\ref{eq:optical_depth}).

The integral in Eq.\,\ref{eq:transmission_spectrum_h} can be split at the radius of reference (because the planet is opaque at all $\lambda$ at smaller radii), and thus Eq.\,\ref{eq:transmission_spectrum_h} becomes
	\begin{eqnarray}
\left[R_{p,0} + h_{eff}(\lambda)\right]^2 & = & R_{p,0}^2 + R_{p,0} c \text{, }\label{eq:transmission_spectrum_h2}\\
 c &\triangleq& 2 \int_{0}^{\infty}  (1+y) (1-e^{-\tau(y,\lambda)}) \text{ d}y, \footnotemark \nonumber \\
 y & = & z/R_{p,0},  \nonumber
\end{eqnarray}
{\footnotetext{Edited from the version published in \textit{Science} to correct Eq.\,\ref{eq:transmission_spectrum_h3} and have consistent definitions of $c$ in Eqs.\,\ref{eq:transmission_spectrum_h2} and \ref{eq:def_yeff}, based on Stephen Messenger's comments.}
leading directly to the expression of the effective atmosphere height
	\begin{eqnarray}
h_{eff}(\lambda)  = R_{p,0} (-1 + \sqrt{1+c}).
\label{eq:transmission_spectrum_h3}
\end{eqnarray}
The embedded atmospheric information can be straightforwardly accessed for most optically active wavelength ranges using
\begin{eqnarray}
 h_{eff}(\lambda) & = & R_{p,0}B(\gamma_{EM}+\ln{A_{\lambda}}),
 \label{eq:h_eff_in_text}
\end{eqnarray}
where $\gamma_{EM}$ is the Euler-Mascheroni constant \cite{Euler1740}---$\gamma_{EM}= \lim_{n \to +\infty} \sum_{k = 1}^n \frac{1}{k} - \ln{n} \approx 0.57722$ (Supplementary Materials Text 1) \cite{science1note5}. In the above equation, $B$ is a multiple of the dimensionless scale height and $A_\lambda$ is an extended slant-path optical depth at reference radius. The exact formulation of $B$ and $A_\lambda$ depends on the extinction processes affecting the transmission spectrum at $\lambda$ (Table\,\ref{tab:A_B}). For Rayleigh scattering, 
\begin{eqnarray}
 B & = & \frac{H}{R_{p,0}} \text{ and }\\
 \label{eq:B_RS_in_text}  
A_{\lambda} & = & \sqrt{2 \pi R_{p,0} H} n_{sc,0} \sigma_{sc}(\lambda),
 \label{eq:Al_RS_in_text}
\end{eqnarray}
where $n_{sc,0}$ and $\sigma_{sc}(\lambda)$ are the number density at $R_{p,0}$ and the cross-section of the scatterers. Conceptually, Eq.\,\ref{eq:h_eff_in_text} tells us the altitude where the atmosphere becomes transparent for a given slant-path optical depth at a radius of reference, $A_\lambda$. For example, if $A_\lambda$ is $10^4$  then the atmosphere becomes transparent at $\approx9$ scale heights above the reference radius.

Most importantly, Eq.\,\ref{eq:h_eff_in_text} reveals the dependency of a transmission spectrum on its key parameters: in particular, $A_\lambda$ is dependent in unique ways on the scale height, the reference pressure, the temperature and the number densities of the main atmospheric absorbents (Supplementary Materials Text\,\ref{app:generalization}), which lead to the mean molecular mass. The uniqueness of these dependencies enables the independent retrieval of each of these key parameters. Therefore, a planet's mass can be constrained uniquely by transmission spectroscopy (Eq.\,\ref{M2}).

\section{\textit{MassSpec}: Applications}
\label{sec:applications}

\subsection{Gas giants}
	With available instruments, \textit{MassSpec} is applicable only to hot Jupiters. Their mean molecular mass is known a priori (H/He-dominated atmosphere: $\mu \approx 2.3$) and their temperature is inferred from, e.g.,  emission spectroscopy. Hence, high SNR/resolution transmission spectra are not required to constrain  their mean molecular mass and temperature independently. Therefore, the measurement of the Rayleigh-scattering slope in transmission is sufficient to yield the mass of a hot Jupiter because it relates directly to its atmospheric scale height. Because $A_\lambda$ depends solely on $\lambda$ through $\sigma_{sc}(\lambda)$ for Rayleigh scattering (Eq.\,\ref{eq:Al_RS_in_text}), by using the scaling law function for the Rayleigh-scattering cross section, $\sigma_{sc}(\lambda) = \sigma_0(\lambda/\lambda_0)^\alpha$, Eq.\,\ref{eq:h_eff_in_text} leads to
\begin{equation}
	\alpha H=\frac{\text{d}R_p(\lambda)}{\text{d}\ln \lambda},
	\label{eq:scaleheightlecavray}
\end{equation} 
with $\alpha =$ -4 \cite{LecavelierDesEtangs2008}. Therefore, using Eqs.\,\ref{M2} and\,\ref{eq:scaleheightlecavray} the planet mass can be derived from 
\begin{equation}
	M_p=-\frac{4kT[R_p(\lambda)]^2}{\mu G\frac{\text{d}R_p(\lambda)}{\text{d}\ln \lambda}}
	\label{eq:M2scattering}.
\end{equation} 
Based on estimates \cite{Madhusudhan2009,Pont2008} of $T\approx$ 1300 K, $\text{d}R_p(\sim 0.8\,\mu\text{m})/\text{d}\ln \lambda\approx$ -920 km, and $R_p(\sim 0.8\,\mu\text{m})\approx$ 1.21$R_{Jup}$ derived from emission and transmission spectra, \textit{MassSpec}'s estimate of HD\,189733b's mass is 1.15 $M_{Jup}$. This is in excellent agreement with the mass derived from RV measurements [1.14$\pm$0.056$M_{Jup}$ \cite{Wright2011}] for this extensively observed Jovian exoplanet. \textit{MassSpec}'s application to gas giants will be particularly important for gas giants whose star's activity prevents a mass measurement with RV [e.g., the hottest known planet, WASP-33b\cite{Collier2010}].

	\subsection{Super-Earths \& Earth-sized Planets}

The pool of planets accessible to \textit{MassSpec} will extend down to super-Earths and Earth-sized planets thanks to the high SNR spectra expected from the \textit{James Webb Space Telescope} (\textit{JWST}; launch date 2018) and the \textit{Exoplanet Characterisation Observatory} (\textit{EChO}; ESA M3 mission candidate). We estimate that with data from \textit{JWST}, \textit{MassSpec} could yield the mass of mini-Neptunes, super-Earths,
and Earth-sized planets up to distances of 500 pc, 100 pc, and 50 pc, respectively, for M9V stars and 200 pc, 40 pc and 20 pc for M1V stars or stars with earlier spectral types (Fig.\,\ref{fig:MassSpec_app_domain_final_in_text};  Supplementary Text\,\ref{app:results}). For \textit{EChO} the numbers would be 250 pc, 50 pc and 13 pc, and 100 pc, 20 pc and 6 pc, respectively.

\begin{figure}[!ht]
 \centering
  \begin{center}
    \includegraphics[trim = 30mm 20mm 40mm 20mm,clip,width=10cm,height=!]{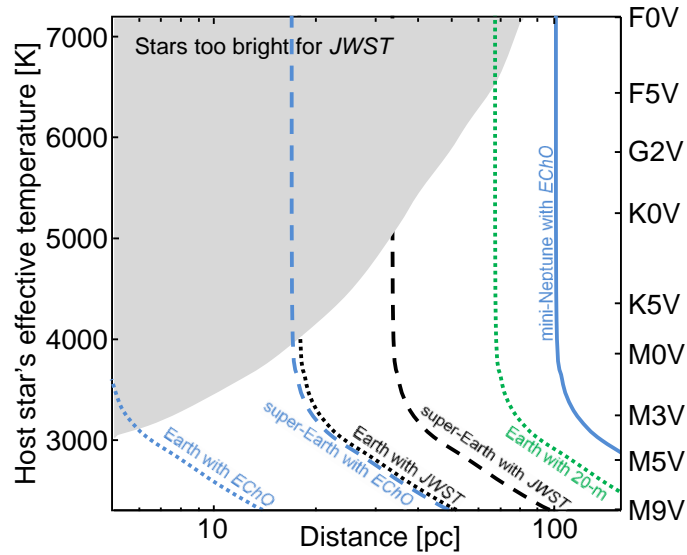}
  \end{center}
  \vspace{-0.7cm}
  \caption{The boundaries of \textit{MassSpec}'s application domain for 200 hours of in-transit observations. Using \textit{JWST}, \textit{MassSpec} could yield the mass of super-Earth and Earth-sized planets up to the distance shown by the black dashed, and dotted lines, respectively. Similarly, the maximum distance to Earth for \textit{MassSpec}'s application based on \textit{EChO}'s observations of a mini-Neptune, a super-Earth, and an Earth-sized planets are shown by the blue solid, dashed, and dotted lines, respectively. The green dotted line refers to the case of an Earth-sized planet observed with a 20-meter space telescope. The grey area show the stars too bright for \textit{JWST}/\textit{NIRSpec} in the R=1000 mode (J-band magnitude $\lesssim$ 7).}
  \vspace{-0.0cm}
  \label{fig:MassSpec_app_domain_final_in_text}
\end{figure}

In particular, if \textit{MassSpec} would be applied to 200 hrs of in-transit observations of super-Earths transiting an M1V star at 15 pc with \textit{JWST}, it would yield mass measurements with a relative uncertainty of $\sim2\%$($\sim10\%$)[$\sim15\%$] for hydrogen(water)[nitrogen]-dominated atmospheres (Figs.\,\ref{fig:mini_neptune_JWST}, \ref{fig:MassSpec_results_in_text_ww}, and \ref{fig:nitrogen_world_JWST}). The larger significance of the mass measurements obtained for hydrogen-dominated super-Earths results from higher SNR of their transmission spectra, which is due to the larger extent of the atmosphere because of the smaller mean molecular mass of H/He. For the same super-Earths with hydrogen(water)-dominated atmospheres, \textit{EChO}'s data should yield mass measurements with a relative uncertainty of $\sim3\%$($\sim25\%$) (Figs.\,\ref{fig:mini_neptune_EChO} and \ref{fig:water_world_EChO}), respectively; for a nitrogen world in the same configuration it will not be possible to constrain the mass (Fig.\,\ref{fig:nitrogen_world_EChO}).

\begin{figure}[!ht]
  \begin{center}
    \includegraphics[trim = 00mm 00mm 00mm 00mm,clip,width=17cm,height=!]{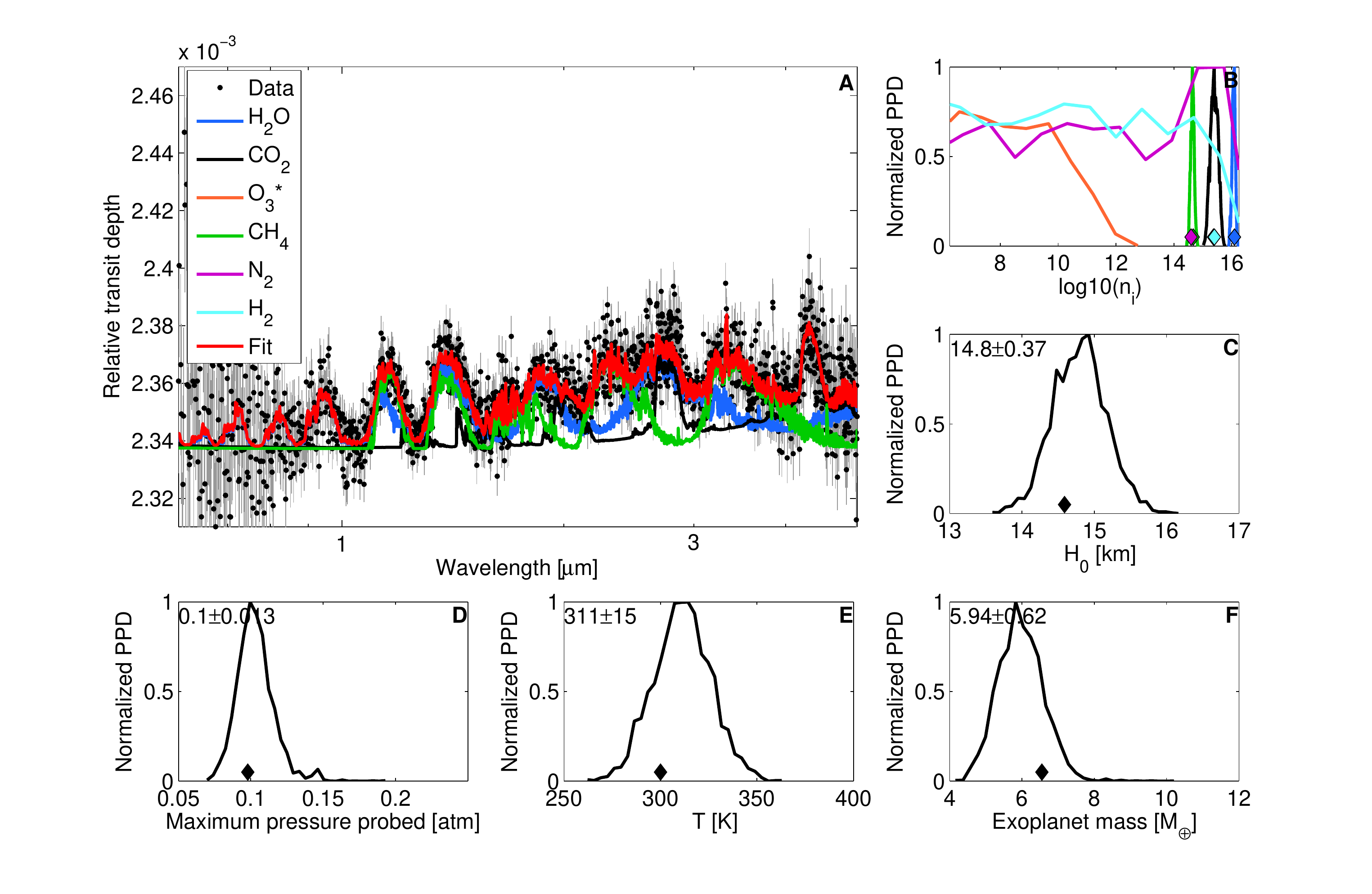}
  \end{center}
  \vspace{-0.6cm}
  \caption{\textit{MassSpec}'s application to the synthetic transmission spectrum of a water-dominated super-Earth transiting a M1V star at 15 pc as observed with \textit{JWST} for a total of 200 hrs in-transit. (\textbf{A}) Synthetic data and the best fit together with the individual contributions of the atmospheric species. (\textbf{B}) Normalized posterior probability distribution (PPD) of the atmospheric species number densities at the reference radius. (\textbf{C}) Normalized PPD for the scale height. (\textbf{D}) Normalized PPD for the pressure at deepest atmospheric level probed by transmission spectroscopy. (\textbf{E}) Normalized PPD for the temperature. (\textbf{F}) Normalized PPD for the exoplanet mass. The diamonds
indicate the values of atmospheric parameters used to simulate the input spectrum and the asterisks in the panel A legend indicate molecules that are not used to simulate the input spectrum.  
The atmospheric properties (number densities, scale height, and temperature) are retrieved with significance yielding to a mass measurement with a relative uncertainty of $\sim10\%$.}
  \label{fig:MassSpec_results_in_text_ww}
\end{figure}

In the future era of 20-meter space telescopes, sufficiently high quality transmission spectra of Earth-sized planets will be available \cite{Kaltenegger2009}. By using \textit{MassSpec}, such facilities could yield the mass of Earth-sized planets transiting a M1V star (or stars with earlier spectral types) at 15 pc with a relative uncertainty of $\sim 5\%$ (Fig.\,\ref{fig:nitrogen_world_20m}). For M9V stars, it would be possible to constrain the mass of Earth-sized planets up to 200 pc and for M1V stars or stars with earlier spectral types, up to 80 pc (Fig.\,\ref{fig:MassSpec_app_domain_final_in_text}).

\begin{figure}[!ht]
  \begin{center}
    \includegraphics[trim = 00mm 00mm 00mm 00mm,clip,width=17cm,height=!]{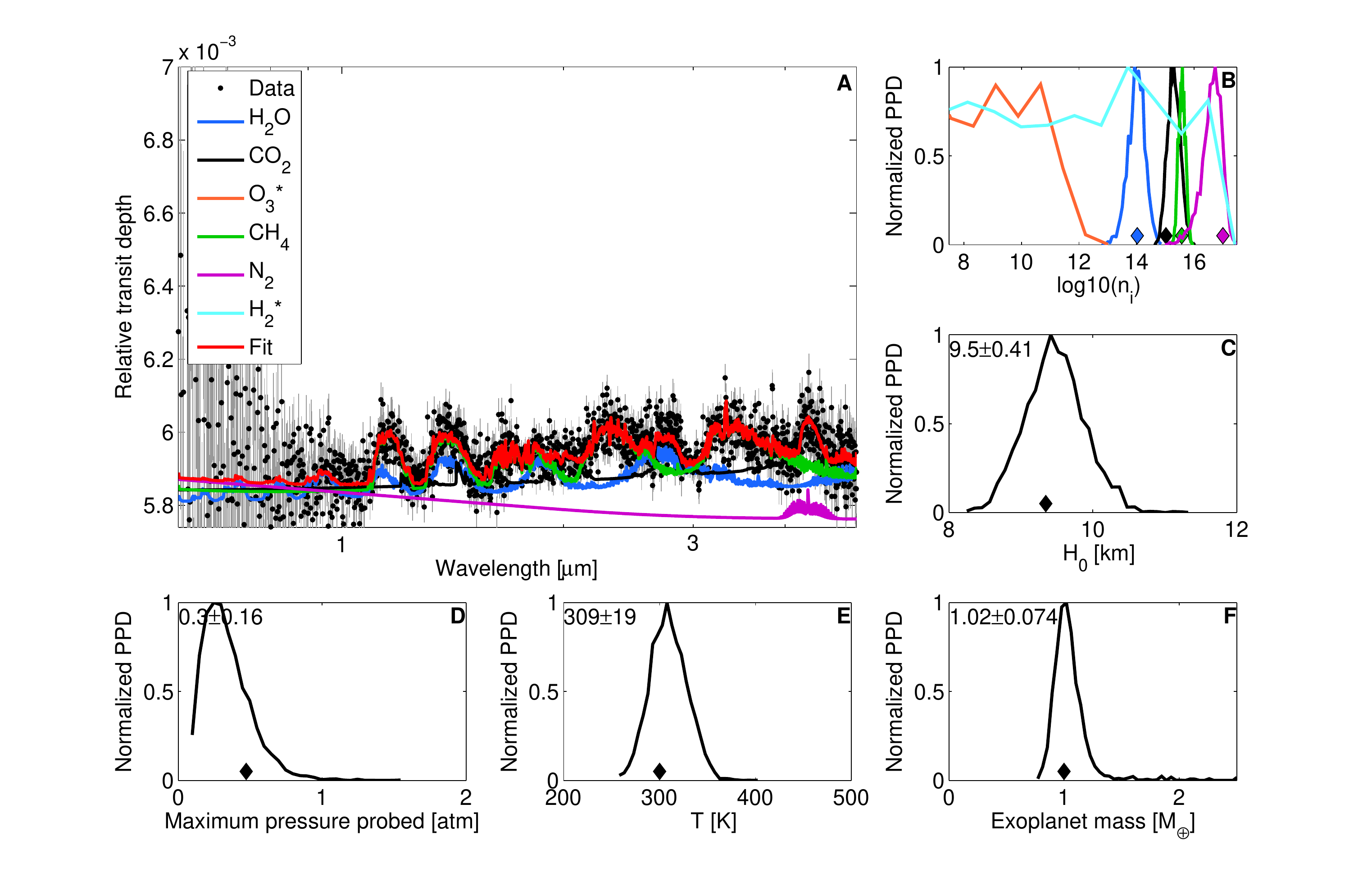}
  \end{center}
  \vspace{-0.6cm}
  \caption{\textit{MassSpec}'s application to the synthetic transmission spectrum of an Earth-like exoplanet transiting a M7V star at 15 pc as observed with \textit{JWST} for a total of 200 hrs in-transit. The panels show the same quantities as on Fig.\,\ref{fig:MassSpec_results_in_text_ww}.  
The atmospheric properties (number densities, scale height, and temperature) are retrieved with significance yielding to a mass measurement with a relative uncertainty of $\sim 8\%$.}
  \label{fig:MassSpec_results_in_text_nw}
\end{figure}

\section{Discussion}
\label{sec:discussion}

\subsection{Habitable Earth-Sized Planets Around Late M dwarfs in the Next Decade}

Late M dwarfs are favorable for any in-transit information such as transmission spectra because of their large ratio of radiance over projected area (Fig.\,\ref{fig:key_ratio_fixed_planet}). For that reason \textit{MassSpec} can be applied to late M dwarfs more distant than other stars, for a given planet (Fig.\,\ref{fig:MassSpec_app_domain_final_in_text}). If they exist, Earth-sized planets may be detected around late M dwarfs before \textit{JWST}'s launch by \textit{SPECULOOS} (\textit{Search for Habitable Planets Eclipsing Ultra-cool Stars}), a European Research Council mission that will begin observing the coolest M dwarfs in 2016. Their mass will not be constrained by RV because of the faintness of their host stars. However, \textit{MassSpec}'s application to their \textit{JWST}'s spectra will yield both their masses and atmospheric properties (Fig.\,\ref{fig:MassSpec_results_in_text_nw}), hence the assessment of their potential habitability. 

\subsection{\textit{JWST}-\textit{EChO} Synergy}

Time prioritization of \textit{JWST} and \textit{EChO} taking into account their synergy would increase the science delivery of both missions. Because the smaller
aperture of \textit{EChO} would enable it to observe brighter stars (i.e., early-type and close-by stars),
\textit{EChO}'s and \textit{JWST}'s time could be respectively prioritized on super-Earths and Earth-sized planets for M9V stars closer than 25 pc and for M1V stars (or stars with earlier spectral type) closer than 10 pc (Fig.\,\ref{fig:MassSpec_app_domain_final_in_text}). Similarly, \textit{EChO}'s and \textit{JWST}'s time could be respectively prioritized on gas giants and super-Earths for M9V stars closer than 125 pc and for M1V stars (or stars with earlier spectral type) closer than 50 pc. \textit{EChO} would be particularly useful to determine the mass---and atmospheric properties---of gas giants because its wide spectral coverage would allow to measure their Rayleigh-scattering slope at short wavelengths.

\subsection{Clouds Will Not Overshadow \textit{MassSpec}}	
	Clouds are known to be present in exoplanet atmospheres \cite{Demory2013} and to affect transmission spectra by limiting the apparent extent of the molecular absorption bands because the atmospheric layers below the cloud deck are not probed by the observations \cite{Barstow2013b}. Therefore, the higher the cloud deck, the larger the error bars are with the  \textit{MassSpec} retrieval method due to the reduced amount of atmospheric information available. For example, the uncertainty on the mass estimate of a water-dominated super-Earth with a cloud deck at 1 mbar is twice the uncertainty obtained for the same planet with a cloud-free atmosphere (Supplementary Text\,\ref{app:results}). However, clouds will not render \textit{MassSpec} ineffectual because they are not expected for pressures below 1 mbar\cite{Howe2012}---which is at least three orders of magnitude (i.e., seven scale heights) deeper than the lowest pressure probed by transmission spectroscopy. In other words, there will always be atmospheric information available from transmission spectroscopy.

\subsection{Complementarity of \textit{MassSpec} and RV}

Transmission spectroscopy is suited for low-density planets and atmospheres and bright or large stars (signal $\propto \rho_p^{-1} \mu^{-1} T_{\star}R_{\star}^{0.5}$), whereas radial velocity measurements are ideal for massive planets and low-mass stars (signal $\propto M_p M_{\star}^{-0.5}$). Therefore, each mass-retrieval method is optimal in a specific region of the planet-star parameter space (Supplementary Text\,\ref{app:scaling_laws}), making both methods complementary.

\subsection{Possible Insights Into Planetary Interiors}
Mass and radius are not always sufficient to obtain insights into a planet's interior. \textit{MassSpec}'s simultaneous constraints on a planet's atmosphere and bulk density may help to break this degeneracy, in some cases. A precision on a planet mass of 3 to 15\%, combined with the planetary radius can yield the planetary average density and hence bulk composition. Even with a relatively low precision of 10 to 15\%, it is possible to infer whether or not a planet is predominantly rocky or predominantly composed of H/He \cite{Seager2007,Fortney2007}. With a higher planet mass precision, large ranges of planetary compositions can be ruled out for high- and low-mass planets, possibly revealing classes of planets with intermediate density to terrestrial-like or ice or giant planets with no solar system counterpart \cite{Rogers2011}. Typically the bulk density alone cannot break the planet interior composition degeneracy, especially for planets of intermediate density. However, measurement of atmospheric species may add enough information to reduce some of the planet interior composition degeneracies---e.g., the rejection of H/He as the dominant species yields constraint on the bulk composition, independently of the mass uncertainty.




\clearpage
\bibliographystyle{Biblio/Science}



\textbf{Acknowledgment}
We are grateful to Andras Zsom and Vlada Stamenkovic for helpful discussions and careful reviews of the manuscript. We also thank Stephen Messenger, William Bains, Nikole Lewis, Brice-Olivier Demory, Nikku Madhusudhan, Amaury Triaud, Michael Gillon, Andrew Collier Cameron, Renyu Hu, and Bjoern Benneke. We thank the anonymous referees who helped to significantly improve the paper. JdW thanks Giuseppe Cataldo and Pierre Ferruit for providing information on \textit{JWST}'s Near Infrared
Spectrograph (\textit{NIRSpec}), and Adrian Belu for further discussions on this matter. JdW acknowledges support from the Wallonie-Bruxelles International, the Belgian American Educational Foundation, and the Grayce B. Kerr Fund in the form of fellowships as well as from the Belgian Senate in the form of the Odissea Prize. JdW is also particularly grateful to the Duesberg-Baily Thil Lorrain Foundation for its support when he conceived this study.



\clearpage

\pagenumbering{roman}
\appendix

\begin{center}
\begin{Huge}
 Supplementary Materials for
\end{Huge}

\begin{Large}
 Constraining Exoplanet Mass from Transmission Spectroscopy
\end{Large}

Julien de Wit and Sara Seager

correspondence to: jdewit@mit.edu

Published 20 December 2013, \textit{Science} \textbf{342}, 1473 (2013)
DOI: \href{http://www.sciencemag.org/content/342/6165/1473}{10.1126/science.1245450}

\end{center}

\textbf{This PDF Files includes:}

 Materials and Methods
 Supplementary Text 
 Figs. S1 to S20
 Tables S1 and S2 
 References \cite{Fortney2005} to (39) 

\clearpage

\renewcommand\thefigure{S.\arabic{figure}}
\setcounter{figure}{0}
\renewcommand\theequation{S.\arabic{equation}}
\setcounter{equation}{0}
\renewcommand\thetable{S.\arabic{table}}
\setcounter{table}{0}
\section{Key parameters of and their effects on a transmission spectrum}
\label{app:dependency}
Here, we derive analytically the dependencies of a transmission spectrum on its main parameters for different extinction processes. In particular, we show that the apparent height takes the form $R_{p,0}B(\gamma_{EM}+\ln{A_{\lambda}})$ for extinction processes such as Rayleigh scattering, collision-induced absorption (CIA), and molecular absorption for most optically active wavelength ranges. $R_{p,0}B$ is a multiple of the scale height and $A_{\lambda}$ is an extended slant-path optical depth at $R_{p,0}$. $B$ and $A_{\lambda}$ summarize how a planet's atmospheric properties are embedded in its transmission spectrum. In particular, the formulation of $B$ and $A_{\lambda}$ reveal the key parameters behind a planet's transmission spectrum. The formulation of $B$ and $A_{\lambda}$ (i.e., the way the atmospheric properties are embedded by transmission spectroscopy) depends on the extinction processes. Therefore, we first approach in details the case of Rayleigh scattering---or any processes with an extinction cross-section independent of $T$ and $p$. Then we extend our demonstration to other processes such as CIA and molecular absorption.

For the coming demonstrations, we will use the following assumptions: (a1) the extent of the optically active atmosphere is small compared to the planetary radius $(z \ll R_p(\lambda))$, (a2) the atmosphere can be assumed isothermal $(d_zT(R_p(\lambda))\simeq0)$, and (a3) the atmosphere can be assumed isocompositional, $d_zX_i(R_p(\lambda))\simeq0$ (where $X_i$ is the mixing ratio of the $i^{th}$ atmospheric constituent). For a later generalization, we specify for each case at which step these assumptions are used.  
	
\subsection{Dependency for Rayleigh scattering}
\label{app:ray}
For extinction processes like Rayleigh scattering, the cross section is independent of pressure and temperature [$\sigma_{\lambda} \neq f_{\lambda}(T,p)$] therefore, using the assumptions a1, a2, and a3, the slant-path optical depth (Eq.\,\ref{eq:optical_depth}) can be formulated as
\begin{eqnarray}
\tau(z,\lambda) & = & \sum_i \sigma_{i}(\lambda) n_{i,0} e^{-z/H} \sqrt{2\pi (R_{p,0}+z) H},
 \label{eq:tau_fortney}
\end{eqnarray}
where the last term comes from the integral over d$x$  \cite{Fortney2005}, or as,
\begin{eqnarray}	
    \tau(y,\lambda) \simeq A_{\lambda} e^{-y/B},
 & \mbox{where } & 
	\left\{{
   \begin{array}{c c c}
     y & = & z/R_{p,0}\\
 A_{\lambda} & = & \sqrt{2 \pi R_{p,0} H} \sum_i n_{i,0} \sigma_{i}(\lambda) \\
 B & = & H/R_{p,0}
  \end{array}}
  	\right.
  	,
  	 \label{eq:tau_var1}
\end{eqnarray}
i.e., $y$ and $B$ are the dimensionless altitude and atmospheric scale height, respectively, and $A_{\lambda}$ is the slant-path optical depth at the reference radius---we recall that the reference radius in any radial distance at which the body is optically thick in limb-looking over all the spectral band of interest. Therefore, Eq.\,\ref{eq:transmission_spectrum_h2} can be rewritten as
\begin{eqnarray}
y_{eff}^2+2y_{eff} & = c = & 2 \int_{0}^{\infty}  (1+y) (1-e^{-A_{\lambda} e^{-y/B}}) \text{ d}y.
\label{eq:def_yeff}
\end{eqnarray}
By solving Eq.,\,\ref{eq:def_yeff}, $y_{eff} = -1 + \sqrt{1+c}$. The integral in Eq.\,\ref{eq:def_yeff} evaluated analytically over $\tau$ is
\begin{eqnarray}
\frac{c}{2} & = & \int_{A_{\lambda}}^{0}  (1-B\ln{\frac{\tau}{A_{\lambda}}}) (1-e^{-\tau}) (-\frac{B}{\tau}) \text{ d}\tau \label{cvA2_0}\\
& = & \int_{A_{\lambda}}^{0} -\frac{B}{\tau} + \frac{B e^{-\tau}}{\tau} + \frac{B^2\ln{\frac{\tau}{A_{\lambda}}}}{\tau} - \frac{B^2\ln{\frac{\tau}{A_{\lambda}}}e^{-\tau}}{\tau} \text{ d}\tau.
 \label{cvA2}
\end{eqnarray}
An evaluation of the integral of each term of Eq.\,\ref{cvA2} leads to
\begin{eqnarray}
\int_{A_{\lambda}}^{0} -\frac{B}{\tau} \text{ d}\tau & = & -B\ln{\tau}|_{A_{\lambda}}^{0}, \label{intA1} \\
\int_{A_{\lambda}}^{0} \frac{B e^{-\tau}}{\tau} \text{ d}\tau & = & B E_i(\tau)|_{A_{\lambda}}^{0}, \label{intA2} \\
\int_{A_{\lambda}}^{0} \frac{B^2\ln{\frac{\tau}{A_{\lambda}}}}{\tau} \text{ d}\tau & = & -B^2\ln{A_{\lambda}}\ln{\tau}|_{A_{\lambda}}^{0} + 0.5B^2\ln^2{\tau}|_{A_{\lambda}}^{0}, \text{ and} \label{intA3} \\
\int_{A_{\lambda}}^{0} - \frac{B^2\ln{\frac{\tau}{A_{\lambda}}}e^{-\tau}}{\tau} \text{ d}\tau & = & B^2\ln{A_{\lambda}}E_i(\tau)|_{A_{\lambda}}^{0} - B^2 \int_{A_{\lambda}}^{0} \frac{\ln{\tau}e^{-\tau}}{\tau} \text{ d}\tau,
 \label{intA4}
\end{eqnarray}
where $E_i(x)$ is the exponential integral. The integral remaining in Eq.\,\ref{intA4} is equal to \newline $ \left[\tau\text{ }_3F_3(1,1,1;2,2,2;-\tau) - 0.5\ln(\tau)(\ln(\tau) + 2\Gamma(0,\tau) + 2\gamma_{EM})\right]|_{A_{\lambda}}^{0}$ where $_pF_q(a_1,...,a_p;b_1,...,b_q;z)$ is the generalized hypergeometric function and $\gamma_{EM}$ is the Euler-Mascheroni constant \cite{Euler1740}---$\gamma_{EM}= \lim_{n \to +\infty} \sum_{k = 1}^n \frac{1}{k} - \ln{n} (\approx 0.57722)$. This insight for the transmission spectrum equations (Eqs.\,\ref{cvA2}-\ref{intA4}) enables further developments of Eq.\,\ref{cvA2_0} using the following series expansions:
\begin{enumerate}
\item $E_i(x)|_{x=0} = \ln{x}+\gamma_{EM}-x+O(x^2)$,
\item $\left[\tau\text{ }_3F_3(1,1,1;2,2,2;-x) - 0.5\ln(x)(\ln(x) + 2\Gamma(0,x) + 2\gamma_{EM})\right]|_{x=0} =\frac{\ln^2{x}}{2}+O(x)$, and 
\item $\left[\tau\text{ }_3F_3(1,1,1;2,2,2;-x) - 0.5\ln(x)(\ln(x) + 2\Gamma(0,x) + 2\gamma_{EM})\right]|_{x\gg1} =\frac{6\gamma_{EM}^2+\pi^2}{12}+O(x^{-5})$.    
\end{enumerate}
The use of this last series expansion is appropriate if $ x \gtrsim 5$, i.e., in optically active spectral bands where $ A_{\lambda} \gtrsim 5$ because absorbers/diffusers affect effectively the light transmission.

In active spectral bands, we can rewrite Eq.\,\ref{cvA2} as 
 \begin{eqnarray}
\frac{c}{2} & = & -B\ln{0} + B\ln{A_{\lambda}} + B(\ln{0} + \gamma_{EM}) - B E_i(A_{\lambda}) - B^2\ln{A_{\lambda}}\ln{0} + B^2\ln^2{A_{\lambda}} + 0.5B^2\ln^2{0}\\&\text{  }& -0.5B^2\ln^2{A_{\lambda}}+B^2\ln{A_{\lambda}}(\ln{0} + \gamma_{EM})-B^2\ln{A_{\lambda}}E_i(A_{\lambda})-0.5B^2\ln^2{0}+B^2\frac{6\gamma_{EM}^2+\pi^2}{12}, \label{simplify_negl_} \\
& = & (\gamma_{EM} + \ln{A_{\lambda}} - E_i(A_{\lambda}))(B+B^2\ln{A_{\lambda}}) + B^2(-\frac{\ln^2{A_{\lambda}}}{2}+ \frac{6\gamma_{EM}^2+\pi^2}{12}) \label{simplify_negl_0}
\end{eqnarray}
Because $ E_i(A_{\lambda}) \ll 1$ and $ \gamma_{EM}\gg B\frac{6\gamma_{EM}^2+\pi^2}{12}$ (recall, $B \ll 1$ because of assumption a1), we can finally write we can rewrite Eq.\,\ref{simplify_negl_0} as
 \begin{eqnarray}
c & = &  (B\ln{A_{\lambda}})^2+2(1+B\gamma_{EM})B\ln{A_{\lambda}}+(2B\gamma_{EM}) .\label{simplify_negl_1}
\end{eqnarray}
 The last step towards the solution to Eq.\,\ref{eq:def_yeff} is to use $B\gamma_{EM} \ll 1$ to write 
 \begin{eqnarray}
1 + c & \simeq &  (B\gamma_{EM} + B\ln{A_{\lambda}} + 1)^2. \label{simplify_negl_c1}
\end{eqnarray} 
 Therefore, we obtain the following solution to Eq.\,\ref{eq:def_yeff}
 \begin{eqnarray}
y_{eff} & \simeq & B(\gamma_{EM}+\ln{A_{\lambda}}) \text{ and} \label{eq:y_eff_ana}\\
\tau(y_{eff}) & \simeq & e^{-\gamma_{EM}} \approx 0.5615
\label{eq:tau_y_eff_ana}
\end{eqnarray}
---note that a first order approximation on $c$ leads to $y_{eff} = B\ln{A_{\lambda}}$ and $\tau(y_{eff}) = 1 $.

Eqs.\,\ref{eq:y_eff_ana} and \,\ref{eq:tau_y_eff_ana} summarize the way a planet's atmospheric properties ($n_{i},T,\mbox{ and }p$) are embedded in its transmission spectrum (Eq.\,\ref{eq:transmission_spectrum_h}). Conceptually, Eq.\,\ref{eq:y_eff_ana} tells us the height (expressed in planetary radius) where the atmosphere becomes transparent. As an example, if $A_\lambda$ is $10^4$ $[\ln(10^4)\approx9]$ then the atmosphere becomes transparent at $\approx9$ scale heights above the reference radius. Eq.\,\ref{eq:tau_y_eff_ana} shows that the slant-path optical depth at the apparent height	is a constant (Fig.\,\ref{fig:transmission_spectrum_basics}, panel C)---this extends previous numerical observations that $\tau_{eq}\approx0.56$ in some case \cite{LecavelierDesEtangs2008}.

The appropriateness of Eqs.\,\ref{eq:y_eff_ana} and\,\ref{eq:tau_y_eff_ana} is emphasized in Fig.\,\ref{fig:Error_on_cst_cross_sect_derivation} that shows the relative deviation on the effective height between numerical integration of Eq.\,\ref{eq:def_yeff} and our analytical solution (Eq.\,\ref{eq:y_eff_ana}). For Earth ($B \approx 0.1\%$), the relative errors in the active spectral bands will be below 0.1\% which corresponds to an uncertainty on $h_{eff}(\lambda)$ (and on $R_p(\lambda)$) below 10 meters---to put this in perspective, recall that the stellar disk and the planet are not resolved.

\subsection{Dependency for collision-induced absorption}
\label{app:cia}
We extend here the results of Section\,\ref{app:ray} to extinction processes with $\sigma\propto P^l $ like CIA. For such processes, Eq.\,\ref{eq:tau_var1} can be rewritten as
\begin{eqnarray}	
    \tau(y,\lambda) \simeq A_{\lambda} e^{-y/B},
 & \mbox{where } & 
	\left\{{
   \begin{array}{c c c}
     y & = & z/R_{p,0}\\
 A_{\lambda} & = & \sqrt{\frac{2 \pi R_{p,0} H}{l+1}} \sum_i \alpha_{l,i,T,0} \\
 B & = & \frac{H}{(l+1)R_{p,0}}
  \end{array}}
  	\right.
  	\mbox{and }
  	 \label{eq:tau_var1cia}
\end{eqnarray}
$\alpha_{l,i,T,0}$ is the temperature- and species-dependent absorption coefficient. For example, for CIA $l = 1$ and the $\alpha_{l,i,T} = K_i(T) n_i^2$ where $K_i(T) = \alpha/n_i^2$ and depends solely on the temperature \cite{Borysow2002}. Now that we obtain the same form for $\tau$ as in Section\,\ref{app:ray} (Eq.\,\ref{eq:tau_var1}) we can apply the same derivation leading to Eqs.\,\ref{eq:y_eff_ana} and\,\ref{eq:tau_y_eff_ana}. Note that $B$ and $A_{\lambda}$ are different than in Section\,\ref{app:ray}---although the general formulation of Eq.\,\ref{eq:tau_var1cia} for processes with $\sigma\propto P^l $ encompasses the case of Rayleigh scattering ($l = 0$).

\subsection{Dependency for molecular absorption}
\label{app:molecules}

We extend here the results of Section\,\ref{app:ray} to molecular absorption. We apply the same strategy as for CIA by showing that the slant-path optical depth (Eq.\,\ref{eq:optical_depth}) can be formulated as $\tau(y,\lambda) = A_\lambda e^{-y/B}$---around $h_{eff}(\lambda)$ and for most $\lambda$---and then relate to the derivation in Section\,\ref{app:ray}. For molecular absorption, the cross-section can be expressed as
\begin{eqnarray}
\sigma_i(\lambda,T,p) = \sum_j S_{i,j}(T) f_{i,j}(\lambda-\lambda_{i,j},T,p),
 \label{lines}
\end{eqnarray}
where $S_{i,j}$ and $f_{i,j}$ are the intensity and the line profile of the $j^{th}$ line of the $i^{th}$ atmospheric species. Each quantity can be approximated by
\begin{eqnarray}
S_{i,j}(T) & \approx & S_{i,j}(T_{ref}) \sum_{m = 0}^{n_T} a_{T,i}T^{m-1} \text{ and} \label{line_intensity_approx} \\
f_{i,j}(\lambda-\lambda_{i,j},T,p) & \approx & A_{i,j}(\lambda,T) \frac{p+a_{i,j}(\lambda,T)}{p^2+b_{i,j}(\lambda,T)}, \label{line_profile_approx}
\end{eqnarray}
where $n_T = 3$ is sufficient to interpolate the line intensity dependency to the temperature \cite{Rothman2009}. $A_{i,j}, a_{i,j},$ and $b_{i,j}$ are the parameters we introduced to model the variation of the line with $p$ at fixed $\{\lambda,T\}$ (Fig.\,\ref{fig:T_p_dependence_of_line_profile}, panel B)---as an example, at low pressure $f_{i,j}(\lambda-\lambda_{i,j},T,p) \approx  A_{i,j}(\lambda,T) a_{i,j}(\lambda,T)/b_{i,j}(\lambda,T)$ where $ A_{i,j}(\lambda,T) a_{i,j}(\lambda,T)/b_{i,j}(\lambda,T)$ is the amplitude of the Doppler profile of at $\{\lambda,T\}$. The second term of Eq.\,\ref{line_profile_approx} is a dimensionless rational function with a zero ($-a_{i,j}$) and a pair of complex conjugate poles ($\pm\sqrt{b_{i,j}}$) (for detail on rational functions and their properties; Section\,\ref{app:TF}). The positions of the zero and the poles in the complex field ($\mathbb{C}$) induce four regimes of specific dependency of $f$ on $T$ and $p$ (Fig.\,\ref{fig:T_p_dependence_of_line_profile}.C)
\begin{enumerate}
\item Doppler regime: while $p < a_{i,j}$, $f_{i,j}$ is independent of $p$ (i.e., $f_{i,j} \approx A_{i,j}a_{i,j}/b_{i,j}$) because neither the zero nor the poles are activated. In terms of distance to the line center ($\nu_j$), the Doppler regime dominates when $(\nu-\nu_j) < \gamma_T$, where $\gamma_T \triangleq \left\lbrace \nu : \mbox{d}^2_{\nu} \ln f_V|_\nu = 0 \right\rbrace$ ($f_V$ and $\gamma_V$ are the Voigt profile and its FWMH, respectively). 

\item Voigt-to-Doppler transition regime: while $p^2 < b_{i,j}$, $f_{i,j} \propto p^1$ because only the zero is activated (i.e., $p \geq a_{i,j}$). In particular, $f_{i,j} = A_{i,j}(p+a_{i,j})/b_{i,j}$. In terms of distance to the line center, this regime dominates when $(\nu-\nu_j) < \gamma_V$. 

\item Voigt regime: while $p^2 \sim b_{i,j}$ and $p \geq a_{i,j}$, $f_{i,j}$ behaves as the rational fraction introduced in Eq.\,\ref{line_profile_approx} because the zero and the poles are activated. In terms of distance to the line center, this regime dominates when $(\nu-\nu_j) \sim \gamma_V$. 

\item Lorentzian regime: while $p^2 \geq b_{i,j}$ and $p \gg a_{i,j}$, $f_{i,j} \propto p^{-1}$ because one zero and two poles are activated. In terms of distance to the line center, this regime dominates when $(\nu-\nu_j) > \gamma_V$. 

\end{enumerate}

While $a_{i,j}$ and $b_{i,j}$ govern the regime of the line profile, $A_{i,j}$ models its variation with temperature; $A_{i,j} \propto T^w$, where $w$ is the broadening exponent---in the Lorentzian regime, the broadening exponent ranges from 0.4 to 0.75 (classical value: 0.5) while it is -0.5 in the Doppler regime \cite{Seager2010}.

Using Eqs.\,\ref{line_intensity_approx} and\,\ref{line_profile_approx} and the ideal gas law, the absorption coefficient can be formulated as
\begin{eqnarray}
\alpha(\lambda,T,p) & = & \sum_i n_i(T,p) \sigma_i(\lambda,T,p) \\
 & = & \frac{Vp}{R} \sum_i X_i \sum_{m = 0}^{3} a_{T,i}T^{m-2}  \sum_j S_{i,j}(T_{ref}) A_{i,j}(\lambda,T) \frac{p+a_{i,j}(\lambda,T)}{p^2+b_{i,j}(\lambda,T)}.
 \label{line_alpha}
 \end{eqnarray}

 For most $\lambda$, the extinction process is dominated by one line. Therefore, the overall dependency of the absorption coefficient can be formulated as
 \begin{eqnarray}
 \alpha(\lambda,T,p) = \Lambda_{\kappa} p \frac{p+a_{\kappa}}{p^2+b_{\kappa}},
 \label{line_alpha2}
\end{eqnarray}
where $\kappa=\left\lbrace\lambda,T,X_i\right\rbrace$ and $\Lambda_{\kappa}, a_{\kappa},$ and $b_{\kappa}$ are model parameters introduced to fit the absorption coefficient variation in the $T-p-X$ space with $\lambda$ being fixed. All of these parameters are known \textit{a priori} from quantum physics and/or lab measurements \cite{Rothman2009}. In particular, for most $\lambda$, $\Lambda_{\kappa}, a_{\kappa},$ and $b_{\kappa}$ are $X_i S_{i,j(\lambda)}(T) A_{i,j(\lambda)}(\lambda)\mbox{, }a_{i,j(\lambda)}(\lambda),$ and $b_{i,j(\lambda)}(\lambda)$, respectively, where $j(\lambda)$ refers to the line that dominates at $\lambda$. We show in Fig.\,\ref{fig:T_p_dependence_of_line_profile}.D that the absorption coefficient behaves like $f_{i,j}$ (Eq.\,\ref{line_profile_approx}) but with an additional zero at $p=0$, which originates from the number density.

Using the assumptions a1, a2, and a3, Eq.\,\ref{eq:optical_depth} becomes
\begin{eqnarray}
\tau(z,\lambda)= 2 \Lambda_{\kappa}\int_{0}^{\infty} \frac{p^2+a_{\kappa}p}{p^2+b_{\kappa}}\text{ d}x,
 & \mbox{where } & 
	\left\{{
   \begin{array}{c c c}
     p& = & p_0 \exp(-z'/H)\\
     z'& = & z'(z,x)\approx \frac{x^2}{2(R_p+z)}+z
  \end{array}}
  	\right.
  	.
 \label{tau_lines_T_cst}
\end{eqnarray}
Transmission spectroscopy probes a limited range of atmospheric layers at each wavelength.
Therefore, only a limited part of the dependency on $T-p$ of the dominant line at $\lambda$ is recorded. We use this property to extend our demonstration assuming that each wavelength records only one regime of dependency (Fig.\,\ref{fig:T_p_dependence_of_line_profile}). By doing so, we show that the slant-path optical depth (Eq.\,\ref{eq:optical_depth}) can be approached by $\tau(y,\lambda) = A_\lambda e^{-y/B}$ in the probed atmospheric layers (i.e., around $h_{eff}(\lambda)$)---and for most $\lambda$. This approach extends the use of Eqs.\,\ref{eq:y_eff_ana} and \,\ref{eq:tau_y_eff_ana} to molecular absorption based on the derivation performed in Section\,\ref{app:ray}.

\subsubsection{Doppler regime}
 When the dominant line at $\lambda$ behaves as a Doppler profile ($p \ll a_{\kappa}$), the absorption coefficient depends on the pressure only through the number density (i.e., similarly to the Rayleigh-scattering case, Section\,\ref{app:ray}). Therefore, Eq.\,\ref{tau_lines_T_cst} can be rewritten as 
\begin{equation}
\tau(z,\lambda) = \sqrt{2 \pi R_{p,0} H} \Lambda_{\kappa} \frac{a_{\kappa}}{b_{\kappa}} p_0 e^{-z/H}.
 \label{tau_lines_T_cst_Dop}
\end{equation}
 
\subsubsection{Voigt-to-Doppler transition regime}

In this regime, $p^2 < b_{\kappa}$, therefore, Eq.\,\ref{tau_lines_T_cst} can be rewritten as 
\begin{equation}
\tau(z,\lambda) = \sqrt{2 \pi R_{p,0} H} \Lambda_{\kappa} \frac{a_{\kappa}}{b_{\kappa}} p_0 e^{-z/H} (1+  \frac{p_0}{\sqrt{2}a_{\kappa}} e^{-z/H}).
 \label{tau_lines_T_cst_V2D_0}
\end{equation}
This regime encompasses the three following subregimes: $p \ll a_{\kappa}$, $p \sim a_{\kappa}$, and $p \gg a_{\kappa}$. The first subregime corresponds to the Doppler regime (Eq.\,\ref{tau_lines_T_cst_Dop}). Eq.\,\ref{tau_lines_T_cst} is rewritten for the second and third subregimes, respectively, as 
 \begin{eqnarray}
 \tau(z,\lambda) & = & \sqrt{\frac{4}{3} \pi R_{p,0} H} \Lambda_{\kappa} \frac{2\sqrt{a_{\kappa}}}{b_{\kappa}} (p_0e^{-z/H})^{3/2}\text{ and}\\
 \label{tau_lines_T_cst_trans}
 \tau(z,\lambda) & = & \sqrt{2 \pi R_{p,0} H} \frac{\Lambda_{\kappa}}{\sqrt{2}b_{\kappa}} (p_0 e^{-z/H})^2.
 \label{tau_lines_T_cst_btw}
\end{eqnarray}

\subsubsection{Voigt regime}

This regime refers to the general formulation of the problem, i.e., when the Doppler and the Lorentzian behaviours affect the line profile with comparable magnitudes. This formulation does not simplify the integral in Eq.\,\ref{tau_lines_T_cst}. Therefore, we rewrite Eq.\,\ref{tau_lines_T_cst} for the transition case $p^2 \sim b_{\kappa} \gg a_{\kappa}^2$
 \begin{eqnarray}
\tau(z,\lambda) & = & \sqrt{2 \pi R_{p,0} H} \Lambda_{\kappa}\frac{2}{\sqrt{b_{\kappa}}} p_0 e^{-z/H}.
 \label{tau_lines_T_cst_infl}
\end{eqnarray}

\subsubsection{Lorentzian regime}

In this regime, $p^2 \geq -b_{i,j}$ and $p \gg -a_{i,j}$; therefore, the absorption coefficient is mostly pressure-independent (Fig.\,\ref{fig:T_p_dependence_of_line_profile}, last panel). As a result, such a regime is not expected to be recorded, under the assumptions a2 and a3.

\subsection{Summary and Discussion}
\label{app:generalization}

Now that we have gone through all the different extinction processes, we can find out what formulation of the effective height is generally true and reveal the key parameters behind a planet's transmission spectrum. We demonstrate that the slant-path optical depth (Eq.\,\ref{eq:optical_depth}) is of the form $\tau(y,\lambda) = A_\lambda e^{-y/B}$, for most $\lambda$. As a result, the apparent atmospheric height can be expressed as $h_{eff} = R_{p,0}B(\gamma_{EM}+\ln{A_{\lambda}})$. We show  the appropriateness of our demonstration in Fig.\,\ref{fig:tau_distribution_isothermal_and_real_Earth} using the numerical simulation of the transmission spectrum of an Earth-sized planet with a isothermal and isocompositional atmosphere---same abundances as at Earth's surface---(details on the transmission spectrum model in Section\,\ref{app:transmission}). It confirms that $\tau_{eq} \approx 0.56$ for a significant fraction of the active bins ($\sim$99$\%$). In addition, we note that a large fraction of the spectral range ($\sim$70$\%$) recorded a $\left\lbrace \propto P^2 \right\rbrace$-dependency of $\tau$. This means that transmission spectroscopy preferably records transitions from the Voigt regime to the Doppler regime. The fact that transmission spectroscopy records preferably the Voigt-to-Doppler transition regime is important because this regime embeds independent information about the pressure. The main reason is that $\gamma_V$ is small, therefore the spectral bins are more likely to be on a line wing, rather than close to the line center. 

Modeling high-resolution spectra based on solving $h_{eff}(\lambda) = [z:\tau(z,\lambda) = e^{-\gamma_{EM}}]$ is adequate, but not correct for all $\lambda$. In particular, we show in Fig.\,\ref{fig:error_on_h_eff_iso_Earth} that for an Earth-sized planet with a isothermal and isocompositional atmosphere---same abundances as at Earth's surface---the error on $h_{eff}(\lambda)$ is below $3\%$ (3$\sigma$) of the scale height. An error on $h_{eff}(\lambda)$ below $3\%$ corresponds to an error of the simulated apparent height below 250 meters---which is sufficient to advocate for the adequacy of modeling transmission spectra based on solving $h_{eff}(\lambda) = [z:\tau(z,\lambda) = e^{-\gamma_{EM}}]$, not its correctness.

\paragraph*{Dependency}
$A_\lambda$ and $B$ record the dependency of a planet's transmission spectrum on atmospheric properties in ways that vary based on the extinction process and the regime effectively recorded at $\lambda$ (Table\,\ref{tab:A_B}). While $B$ is solely affected by $H$, $A_\lambda$ is affected in independent ways by the following parameters:
\begin{itemize}
\item The scale height, $H$, affects $A_\lambda$ through $\sqrt{H}$ for all extinction processes---geometry factor from the light path.

\item The number densities of the main atmospheric absorbents, $n_i$, affects $A_\lambda$ proportionally for molecular absorption (recall $\Lambda_{\kappa}p_0 = n_{i,0} S_{i,j(\lambda)}(T) A_{i,j(\lambda)}(\lambda)$) and for Rayleigh scattering, while their square affect $A_\lambda$ for CIA---resulting in stronger constraints on $n_i$ when assessed primarily from CIA signal.

\item The reference pressure, $p_0$, affects $A_\lambda$ for molecular absorption in the Voigt-to-Doppler regime ($A_\lambda \propto p_0^{0.5}$ if $p \sim a_{\kappa}$ and $A_\lambda \propto p_0^1$ if $p \gg a_{\kappa}$), which is recorded by most of the spectral bins.

\item The temperature, $T$, affects $A_\lambda$ in two ways: (i) through $\Lambda_{\kappa}$ due to the line-intensity dependence on $T$ (as a third-order polynomial) and (ii) through the line-profile dependence on $T$ ($T^w$-dependency of the line broadening). In addition, processes like CIA are also dependent on $T$, in a way that is known a priori from quantum mechanics \cite{Borysow2002}. 

\end{itemize}
Although both the information on the temperature and the number densities are embedded in $\Lambda_{\kappa}$, $T$ and $n_i$ affect $\Lambda_{\kappa}$ in specific/unique way. $\Lambda_{\kappa} \propto n_i$ independently of $\lambda$, while it depends on $T$ in a way that varies with $\lambda$ through the line intensity and profile---dependency known a priori from quantum mechanics or lab measurements.

In summary, the key parameters of a transmission spectrum---the atmospheric scale height, the number densities of the main absorbents, the temperature, and the pressure---affect it independently. Therefore, it is theoretically possible to retrieve independently each of these parameters, although they are embedded in the spectrum.

\begin{table}[!h]
\caption{Dependency of $A_{\lambda}$ and $B$ to the regimes recorded \label{tab:A_B}}
	\centering
	\setlength{\extrarowheight}{-0pt}
	\setlength{\tabcolsep}{2pt}

	\begin{tabular}{c|r l|c}
	
	\hline\hline
	\textbf{Extinction process}---Recorded regime & \multicolumn{2}{c}{\textbf{$A_{\lambda}$}} & \textbf{$B$}\\
	\hline
	{\textbf{Rayleigh scattering}} &  $\sqrt{2 \pi R_{p,0} H} $&$ \sum_i n_{i,0} \sigma_{i}(\lambda)$ & $\frac{H}{R_{p,0}}$\\
	{\textbf{Collision Induced Absorption}} &  $ \sqrt{\text{ } \pi R_{p,0} H} $&$ \sum_i K_i(T) n_{i,0}^2$ & $\frac{H}{2R_{p,0}} $\\
	\hline
	{\textbf{Molecular absorption}} & & \\
	{Doppler regime} ($p < a_{\kappa}$) & $\sqrt{2 \pi R_{p,0} H} $&$ \Lambda_{\kappa} \frac{a_{\kappa}}{b_{\kappa}} p_0$ & $\frac{H}{R_{p,0}}$\\
	{Voigt-to-Doppler transition regime} ($p^2 < b_{\kappa}$)  & & \\
	 {{if $p \sim a_{\kappa}$}} & $\sqrt{\frac{4}{3} \pi R_{p,0} H} $&$ \Lambda_{\kappa} \frac{2\sqrt{a_{\kappa}}}{b_{\kappa}} p_0^{1.5} $ & $\frac{H}{1.5R_{p,0}}$ \\
	  {{if $p \gg a_{\kappa}$}} & $\sqrt{\text{ } \pi R_{p,0} H} $&$ \Lambda_{\kappa} \frac{1}{b_{\kappa}} p_0^{2} $ & $\frac{H}{2R_{p,0}}$ \\
	{Voigt regime} & & \\
	  {{if $p^2 \sim b_{\kappa} \gg a_{\kappa}^2$}} & $\sqrt{2 \pi R_{p,0} H} $&$ \Lambda_{\kappa} \frac{2}{\sqrt{b_{\kappa}}} p_0 $ & $\frac{H}{R_{p,0}}$ \\

	\hline

	\end{tabular}
	
\end{table}

\paragraph*{Discussion}
The analytical derivations performed in this section required using the three following assumptions: (a1) the extent of the optically active atmosphere is small compared to the planetary radius ($z \ll R_p$), (a2) the atmosphere can be assumed isothermal $[d_zT(R_p(\lambda))\simeq0]$, and (a3) the atmosphere can be assumed isocompositional $[d_zX_i(R_p(\lambda))\simeq0]$. We discuss below how Section\,\ref{app:dependency}'s conclusions are mostly unaffected by the relaxation of the assumptions a2 and a3---a1 being justified. In particular, we explain conceptually that the effect of each key parameter remains unique while relaxing these assumptions, meaning that \textit{MassSpec} can be applied to any exoplanet atmosphere, theoretically.

The analytical derivations become more complex if assumptions a2 and a3 are relaxed, therefore we cannot extend the derivations  for the general case. Yet, it is possible to describe conceptually how these relaxations affect the results of this section. 

(i) A non-isothermal atmosphere translates primarily into an altitude-dependent scale height [$H=H(z)$]---therefore, it is required to model the planet's atmosphere with a $z$-dependent scale height in the retrieval method (Section\,\ref{app:numerics}). However, the temperature can still be self-consistently retrieved from a planet's transmission spectrum because it affects specifically the slant-path optical depth profile $[\tau(r,\lambda)]$ through the extinction cross section profile [$\sigma_{i}(T(r'),p(r'),\lambda)$].
For molecular absorption, $T$ affects $\sigma_{i}(T(r'),p(r'),\lambda)$ through the line intensities and profiles in ways that are known a priori from quantum mechanics and/or lab measurements. $T$ affects mainly the line intensities through the total internal partition sum (TIPS) and the Boltzmann populations for molecular absorption. The TIPS describes the overall population of the molecule's quantum states, and is solely dependent on the molecular structure (i.e., the species) and the local temperature. The effect of temperature on the TIPS can be appropriately approximated by a third order polynomial. On the other hand, the temperature effect on the population of a molecule's individual state depends solely on the temperature and the energy of the state. Therefore, the extinction cross-section depends on the temperature in ways that vary with the wavelength, in opposition to the TIPS. Finally, the temperature affects the line broadening as $\propto T^w$, where $w$ is the broadening exponent and ranges from 0.4 to 0.75 in the Lorentzian regime (classical value: 0.5),  while it is -0.5 in the Doppler regime \cite{Seager2010}.

The overall effect of a change in $T$ cannot be compensated/mimicked by other atmospheric parameters because it is specific---and a priori known. For example, although a local decrease in temperature could be compensated at the zeroth order by an increase of the local number densities:
\begin{itemize}
\item The increase in number densities required to compensate the change in the line intensities will be inconsistent with the increase required to mitigate the change in local pressure. While a molecule's local number density will have to scale as a third order polynomial to compensate line intensity changes, it should scale as $1/T$ to compensate the change in pressure.
\item The compensation enabled by an increase in number densities is wavelength-independent while the effect of $T$ on the absorption coefficient is strongly wavelength-dependent.
\end{itemize}

(ii) A non-isocompositional atmosphere translates into different number density scale height for each component. Alike in the case of temperature changes, changes in composition affect the slant-path in specific ways, preserving \textit{MassSpec}'s capability for obtaining independent constraints on the parameters of the mass equation (Eq.\,\ref{M2}). A change in number density is species specific and, hence, cannot be compensated by a change in temperature, scale height or reference pressure. 

In some case, molecules such as water may require the use of an individual scale height for proper retrieval. The use of individual scale heights will be of primary importance for planets in habitable zone as a significantly smaller scale height for the water number density could indicate the presence of water surface reservoir (a key to assessing habitability). For such applications, comparisons with the scale height of other molecules will be required. Molecules that are expected to present constant mixing ratio throughout the atmosphere could then be considered as independent markers to extract the pressure scale height. The best marker candidate is carbon dioxide because (i) it is a molecule chemically stable that is thus well-mixed in a planet's atmosphere and (ii) it presents numerous strong absorption bands that enable CO$_2$ detection at low abundance (down to $\sim0.1$ppm). In the extreme cases where water and carbon dioxide are depleted from a planet's atmosphere due to condensation, nitrogen and/or hydrogen---which are known to be chemically stable at temperatures leading to CO$_2$ condensation---would be the dominant species and, hence, the primary marker for the pressure scale.

\paragraph*{Modeling transmission spectra by solving $h_{eff}(\lambda) = [z:\tau(z,\lambda) = e^{-\gamma_{EM}}]$}

As shown in Section\,\ref{app:ray}, the apparent height is of the form $R_{p,0}B(\gamma_{EM}+\ln{A_{\lambda}})$ when $\tau_{eq}\approx0.56$. We show in Fig.\,\ref{fig:tau_eq_vs_a_lambda_Earth} that for $50\%$ of the optically active bins, our formulation for $h_{eff}$ is still adequate for Earth, i.e., for a planet with atmospheric temperature and mixing ratios strongly dependent on the altitude. We emphasize in Fig.\,\ref{fig:tau_eq_vs_a_lambda_Earth_without_water} that water is the main origin for the deviation of $\tau_{eq}$'s distribution from $\sim0.56$ shown in Fig.\,\ref{fig:tau_eq_vs_a_lambda_Earth}. The removal of water from Earth's atmosphere leads to $\tau_{eq} \approx 0.56$ for 80\% of the active bins. The effect of water on $\tau_{eq}$'s distribution is due to the strong variation of the water mixing ratio with the altitude close to Earth's surface. The strong variation of the water mixing ratio with the altitude invalidates assumption a3 and therefore the demonstration in Section\,\ref{app:molecules}, meaning that $h_{eff}$ deviates from the formulation $R_{p,0}B(\gamma_{EM}+\ln{A_{\lambda}})$. As we discussed above, it is not because $h_{eff}$ deviates from $R_{p,0}B(\gamma_{EM}+\ln{A_{\lambda}})$ that the specificity of the dependency of a transmission spectrum to its parameters is invalid too---we are just unable to provide the general derivation in such a general atmosphere. Note that the significant effect of a specific species's scale height on a transmission spectrum is favorable for hability assessment (see previous paragraph).

We show in Fig.\,\ref{fig:error_on_h_eff_Earth} the error on $h_{eff}(\lambda)$ when modeling Earth's transmission spectra using $h_{eff}(\lambda) = [z:\tau(z,\lambda) = e^{-\gamma_{EM}}]$. The error on $h_{eff}(\lambda)$ is below $18\%$ (3$\sigma$). An error on $h_{eff}(\lambda)$ below $18\%$ corresponds to an error of the simulated apparent height below 1500 meters---which is sufficient to advocate for the adequacy of modeling transmission spectra based on solving $h_{eff}(\lambda) = [z:\tau(z,\lambda) = e^{-\gamma_{EM}}]$, not its complete accuracy.

Finally, note that solving $h_{eff}(\lambda) = [z:\tau(z,\lambda) = e^{-\gamma_{EM}}]$ is computationally much more efficient than a direct numerical integration of Eq.\,\ref{eq:transmission_spectrum_h}. Solving $h_{eff}(\lambda) = [z:\tau(z,\lambda) = e^{-\gamma_{EM}}]$ solely requires knowing the parameters $\Lambda_{\kappa}, a_{\kappa},$ and $b_{\kappa}$ (Eq.\,\ref{line_alpha2})---i.e., the dependency of each species' extinction cross-section on the pressure and the temperature---from quantum physics and/or lab measurements \cite{Rothman2009}.

\clearpage

\section{\textit{MassSpec}'s potential using future facilities}
\label{app:numerics}

In this second section we introduce the methods developed to explore what constraints \textit{MassSpec} can enable on an exoplanet's mass and atmosphere from future high-SNR transmission spectra. Therefore, we \textbf{(1)} generate theoretical transmission spectra, \textbf{(2)} estimate their instrumental output and \textbf{(3)} perform their analysis. We first present our transmission model. Then we describe our noise model for \textit{JWST/NIRSpec}, \textit{EChO}, and a synthetic 20-meter telescope. We introduce our atmospheric retrieval method and the synthetic exoplanet scenarios used for the assessment of \textit{MassSpec}'s potential. Finally, we present and discuss our retrieval method results.

\subsection{Transmission Spectrum Model}
\label{app:transmission}
	We model high resolution\footnote{It is necessary to model the radiative transfer process at high resolution to approach adequately the absorption lines and their effects on the stellar light. Once the simulated transmission spectrum is modeled it can be binned down to the facility's spectral resolution for comparison.} ($R = \lambda/\Delta \lambda = 10^5$) transmission spectra following Eq.\,\ref{eq:transmission_spectrum_h}: (i) we calculate the slant-path optical depth for different altitudes $(\tau(z,\lambda))$ and (ii) integrated the contribution of each projected atmospheric annulus to the overall flux drop---$2\pi r \mbox{d}r (1-\exp[-\tau(r,\lambda)]) $ (Fig.\,\ref{fig:transmission_spectrum_basics}). The extinction cross section accounts for molecular absorption, collision-induced absorption (CIA), and Rayleigh scattering. We computed the monochromatic molecular absorption cross sections from the HITRAN 2008 database \cite{Rothman2009} and approximate the Voigt line profile according to \cite{Liu2001} in order to increase the computational speed. We use opacity tables from \cite{Borysow2002} for H$_2-$H$_2$ CIA. We determine the Rayleigh-scattering cross section, $\sigma_{R,i}$ in cgs units, from
	\begin{equation}
\sigma_{R,i}(\lambda) = \frac{24\pi ^3}{n_{i}^2\lambda^{4}} \sqrt{\frac{N_{i}(\lambda) ^2-1}{N_{i}(\lambda) ^2+2}} F_{i}(\lambda),
\label{eq:rayleight}
\end{equation} 
	where $\lambda$ is the wavelength and $n_{i}$, $N_{i}(\lambda)$, and $F_{i}(\lambda)$ are the number density, the refractive index, and the King correction factor for the depolarization of the i$^{th}$ atmospheric species. In particular, we use the refractive indices of N$_2$, CO$_2$, CO, CH$_4$, N$_2$O from \cite{Sneep2005}\footnote{We observe discrepancies between the measured data and the functional forms proposed in \cite{Sneep2005} for the refractive indices of CO$_2$ and CO---their equations (13) and (17). Therefore, we use the following corrected forms:
	\begin{equation}
\begin{split}
\frac{n_{CO_2}-1}{1.1427 \times 10^{\boldsymbol{3}}}  &= \frac{5799.25}{(128908.9)^2-\lambda^{-2}} + \frac{120.05}{(89223.8)^2-\lambda^{-2}}\\ 
&+ \frac{5.3334}{(75037.5)^2-\lambda^{-2}} + \frac{4.3244}{(67837.7)^2-\lambda^{-2}}+\frac{\boldsymbol{0.1218145}}{(2418.136)^2-\lambda^{-2}}, 
\end{split}
\label{eq:n_sneepco2}
\end{equation}
	\begin{equation}
\frac{n_{CO}-1}{1 \times 10^{8}} = 22851 + \frac{0.456\times 10^{\boldsymbol{14}}}{(71427)^2-\lambda^{-2}}.
\label{eq:n_sneepco}
\end{equation}
	}.

	A transmission spectrum simulation requires the computation of millions of absorption lines for each atmospheric species and numerous $T-p$ conditions. In addition, we develop our retrieval method in a Bayesian framework which requires a large number of transmission model runs to converge. Therefore, we determine the extinction cross section for each component of HITRAN as a function of $\lambda$, $T$ and $p$ to interpolate later for the required conditions. In particular, we generate the extinction cross section 4-D array 
	\begin{itemize}
	 \item at a spectral resolution of $10^5$ for 0.4$\mu$m to 250$\mu$m,
	 \item for 17 pressure values spread in the $\log_{10}p[\text{Pa}]$ space from 7 to -3 with a higher density around 4---because most of the information is recorded around 1 mbar,
	 \item for 12 temperature values homogeneously spread from 150 K to 700 K.
	 \end{itemize} 
	 
	 We validate our extinction cross section model with \cite{Benneke2012}. In addition, we validate our transmission spectrum model comparing a synthetic Earth transmission spectrum with \cite{Kaltenegger2009} and \cite{Hu2013}. Except for a 3.2$\mu$m-water signature\footnote{We observed an additional water band at 3.2$\mu$m in \cite{Kaltenegger2009}; it could be due to a complementary list used in this study. The HITRAN water absorption lines are uniformly spaced in terms of wavenumber, as expected from quantum mechanics. Furthermore, HITRAN's aim is to provide the spectroscopic parameters required to simulate the transmission and emission light for Earth-like conditions. Therefore, it is unlikely that such a significant water band would not be included in HITRAN.}, our simulation is consistent with our references; for that reason, we consider our transmission model appropriate for the present study.

\subsection{Facilities' Noise Model}

\subsubsection{\textit{JWST/NIRSpec}}
\label{app:nirspec}
	\textit{NIRSpec} is the Near-Infrared Spectrograph for the \textit{James Webb Space Telescope} (\textit{JWST}). The purpose of \textit{NIRSpec} is to provide low (R = 100), medium (R = 1000), and high-resolution (R = 2700) spectroscopic observations from 0.6 to 5 $\mu$m. In the present study we focus on the medium resolution mode because \textit{MassSpec} requires a sufficient spectral resolution. Furthermore, its larger spectral dispersion enables the observation of brighter stars. Note that the effective observation time is then reduced by a factor of three because of its three grisms---alike for the high resolution mode. 
	
	In order to estimate the noise budget of a telescope instrument one needs to \textbf{(1)} evaluate its total optical throughput\footnote{The total optical throughput of a telescope is the ratio of photons collected by its primary mirror to electrons read over a spectral resolution element of its detector.} and \textbf{(2)} derive the instrument duty-cycle for a given target.	

\paragraph*{\textit{JWST/NIRSpec} Total Optical Throughput}
	 We determine the total optical throughput of \textit{JWST/NIRSpec} following the procedure described in \cite{Boker2010} and model the detector pixel efficiency as a plateau at the 75\% level with a drop of 3\% at the pixel edges and integrate the spectrograph $\lambda$-dependent PSF over the pixel grid to estimate the resolution element sensitivity---accounting for distortions for wavelengths under 1 $\mu$m using the 1 $\mu$m PSF-size \cite{Boker2010}.

	We present the throughput budget summary for the \textit{NIRSpec} medium-resolution mode in Fig.\,\ref{fig:nirspec_perfo}. Fig.\,\ref{fig:nirspec_perfo} also shows our estimate of the flux fraction going to the brightest pixel of each resolution element which we will use further to derive the saturation time of the detector. 
		
	\paragraph*{JWST/NIRSpec Noise Budget}
	
	We derive the budget noise from the instrumental performance shown in Fig.\,\ref{fig:nirspec_perfo}. First, we estimate the electron flux produced on the brightest pixel of the detector for this given star, using the data of Fig.\,\ref{fig:nirspec_perfo}. Then, we derive the saturation time from the electron flux and a pixel well capacity of 60,000 e$^-$ to ensure linear response. We derive the variance of a frame, $\sigma_{1F}(\lambda)$, including a readout noise of 6 e$^-$\,px$^{-1}$ rms and a dark current of 0.03 e$^-$\,(px\,s)$^{-1}$ and assuming the read mode to be MULTIACCUM-2x1 \cite{Boker2010}. We derive the duty cycle based on a readtime of $\sim$0.53 sec, which, together with $\sigma_{1F}(\lambda)$, leads to the signal-to-noise ratio (SNR) for a given observation time, $t$.

	\paragraph*{\textit{JWST/NIRSpec} Noise Model Comparison}

	Unlike \cite{Deming2009}, we consider \textit{JWST/NIRSpec} observations will be photo-noise dominated; the pixel-phase of \textit{NIRSpec}'s detector is expected to be negligible. In particular, the pixel-phase of \textit{NIRSpec}'s detectors will mainly deviate from a plateau at the pixel edge with a relative drop below 2\% induced mainly by cross-talk (i.e., the information is mainly transferred, not lost). These variations are more than an order of magnitude less than presented in \cite{Deming2009}. Furthermore, even for current facilities significantly affected by pixel-phase like the \textit{Spitzer Space Telescope}, the observations are within 10 to 20\% of the photon noise after systematic correction. In addition, we take into account the necessity to change the grisms to obtain a full spectrum \cite{Boker2010}. For practical purposes including stability and baseline follow-up, we consider that one transit is observed in a unique grism. For that reason we further scaled the SNR by $1/\sqrt{3}$ to include the inherent sharing of the integration time between the three channels of \textit{NIRSpec}'s medium-resolution module.
	
\subsubsection{\textit{EChO}}

We use \textit{EChO}'s noise model introduced in \cite{Barstow2013}. In particular, we use a telescope effective area of $1.13$ square meter, a detector quantum efficiency of $0.7$, a duty-cycle of $0.8$, and an optical throughput of 0.191 from 0.4 to 0.8 $\mu$m, 0.284 from 0.8 to 1.5
$\mu$m, 0.278 from 1.5 to 2.5 $\mu$m, 0.378 from 2.5  to 5 $\mu$m, 0.418 from 5 to 8.5 $\mu$m,
0.418 from 8.5 to 11 $\mu$m, 0.326 from 11 to 16 $\mu$m.

\subsubsection{Future-generation telescope}

We use a scaled-up version of \textit{EChO} to model the 20-meter telescope. In particular, we use a spectral resolution of $1000$, a detector quantum efficiency of $0.7$, a duty-cycle of $0.8$, and an optical throughput of 0.4 from 0.4 to 16 $\mu$m.

\subsection{Atmospheric Retrieval Method}
\label{app:retrieval}
We use an adaptive Markov Chain Monte Carlo (MCMC) algorithm to retrieve the properties of an exoplanet's atmosphere embedded in its transmission spectrum (Eq.\,\ref{eq:transmission_spectrum_h}). The key improvement compared to previous studies \cite{Madhusudhan2009,Benneke2012} is the use of a self-consistent set of parameters derived from the first principle of transmission spectroscopy (S\ref{app:dependency}) that uniquely constrains the planetary mass. In other words, we use as jump parameters\footnote{Jump parameters are the model parameters that are randomly perturbed at each step of the MCMC method.}: the temperature, the pressure scale height and the species number density at a given planetary radius---whose choice does not affect the retrieval method, similarly to the ``reference radius''.    We assume a uniform prior distribution for all these jump parameters and draw at each step a random stellar radius based on a Gaussian prior assuming a 1$\%$ relative uncertainty on the stellar radius.

\subsubsection{Synthetic exoplanet scenarios}
\label{app:scenarios}
	We first assess \textit{MassSpec}'s capabilities for different super-Earths observed with \textit{JWST/NIRSpec} and \textit{EChO}. We use GJ\,1214b's sizes \cite{Charbonneau2009} and the ``hot mini-Neptune'', ``Hot Halley world'', and ``nitrogen-rich world'' scenarios introduced in \cite{Benneke2012} to provide a representative overview of \textit{MassSpec}'s potential. Then, we investigate the effects of clouds on \textit{MassSpec}'s capabilities using the water-dominated super-Earth and three cloud-deck scenarios at 100, 10, and 1 mbar. Note that 100 mbar corresponds to Earth's and Venus' clouds's pressure level and 1 mbar is the lowest pressure where thick clouds are expected \cite{Howe2012}. Finally, we test \textit{MassSpec}'s capabilities for an Earth-sized planet observed (1) with \textit{JWST} and (2) with a future-generation 20-meter space telescope. We highlight that \textit{MassSpec}'s application to Earth-sized planet with \textit{JWST} (and \textit{EChO}) requires taking advantage of late M dwarfs (Section\,\ref{app:scaling_laws_snr}). For this last application to Earth-like planets, we choose to focus on a nitrogen-world (``Earth-like" planet) because (1) the first exoplanet to be confirmed habitable is likely to be a hot desert world \cite{Zsom2013}, (2) in most cases hydrogen should have escaped from an Earth-sized planet atmosphere, and (3) nitrogen-dominated atmospheres are less favorable for transmission spectroscopy due to their larger mean molecular mass (Eq.\,\ref{eq:SNRt_general_scalinglaw}) hence \textit{MassSpec}'s capabilities derived from those are more conservative.
	
	We assume a total in-transit observation time of 200 hrs \cite{Deming2009} and a M1V host star located at 15 pc, except for the Earth-like planet observed with \textit{JWST} for 200 hrs around a M7V at 15 pc. We use temperature profiles similar to \cite{Miller-Ricci2009}'s temperature (in particular, $T(p\gtrsim1\text{ mbar})\approx300$ K) and assume a well-mixed atmospheres.

\subsection{Results \& Discussion}
\label{app:results}

We find that future space based facilities designed for exoplanet atmosphere characterization will also be capable of mass measurements for super-Earths and Earth-sized planets with a relative uncertainty as low as $\sim2\%$---a precision that has not yet been reached using RV measurements, even for the most favorable cases of hot Jupiters.

\paragraph{Super-Earths with \textit{JWST} and \textit{EChO}}

See Main Text (Section \textit{MassSpec}'s Applications). 

\paragraph*{The effect of clouds on \textit{MassSpec}'s}	
	Clouds are known to be present in exoplanet atmospheres \cite{Demory2013} and to affect transmission spectra by limiting the apparent extent of the molecular absorption bands as the atmospheric layers below the cloud deck are not probed \cite{Barstow2013b}. However, \textit{MassSpec} is not rendered ineffectual by clouds because clouds are deeper than the lowest pressure probed by transmission spectroscopy. In other words, spectral features are still detectable in the presence of clouds but with a reduced significance (compare panels A of Figs.\,\ref{fig:water_world_JWST_C1}-\ref{fig:water_world_JWST_C3}): the higher the cloud deck, the larger is the uncertainty on the mass estimate due to the reduced amount of atmospheric information available. We show the effect of cloud decks at 100, 10, and 1 mbar in Figs.\,\ref{fig:water_world_JWST_C1}, \ref{fig:water_world_JWST_C2}, and \ref{fig:water_world_JWST_C3}, respectively. Cloud decks at 100 and 10 mbar affect marginally \textit{MassSpec}'s capabilities because a limited fraction of the spectral bins probe deeper than 10 mbar in transmission. However 1 mbar cloud deck affects significantly atmospheric retrieval results, increasing the uncertainty on the mass estimate by a factor of $\sim2$ over the one derived from the cloud free scenario. The increased uncertainty in the mass estimate results mainly from an increased uncertainty on the scale height and the mean molecular mass estimates. The reason is that (1) the significance of scale-height estimates are reduced because the signatures of atmospheric species are truncated and (2) additional atmospheric scenarios are now possible (e.g., hydrogen-dominated) because numerous combination of molecular signatures could be masked by a high-cloud deck.

\paragraph*{Pushing down to Earth-sized planets within the next decade}

We show that \textit{MassSpec} could yield the mass of Earth-sized planets transiting late M dwarfs within 15 pc of Earth with a relative uncertainty of $\leq8\%$ using \textit{JWST} (Fig.\,\ref{fig:MassSpec_results_in_text_nw}). This reemphasizes the key role of late M dwarfs for transmission spectroscopy as the first Earth-sized planets to be thoroughly characterized will be transiting such stars (Fig.\,\ref{fig:key_ratio_fixed_planet}). It is therefore possible that the first Earth-sized planet to be confirmed habitable would be one of these---potentially a hot desert world \cite{Zsom2013}. That is why, \textit{SPECULOOS} is of particular interest for the short- and medium-term future for aiming to detect Earth-sized planets around nearby late M dwarfs. Finally, for Earth-sized planets transiting M1V stars or stars with earlier spectral types within 15 pc of Earth, we find that a future-generation 20-meter space telescope will yield their masses with a relative precision $\leq5\%$ (Fig.\,\ref{fig:nitrogen_world_20m}). 

\paragraph{Gas Giants}

\textit{JWST}'s and \textit{EChO} 's observation of planets larger than super-Earths will yield to high-significance mass measurements with a limited number of transits. We estimate that for gas giants half a dozen of transits can yield mass measurements with a relative uncertainty $\leq 5\%$, using the scaling law (Eq.\,\ref{eq:SNRt_general_scalinglaw}). \textit{EChO} would be particularly useful to determine the mass---and atmospheric properties---of gas giants because its wide spectral coverage would allow to measure their Rayleigh-scattering slope at short wavelengths. Independent high-significance mass measurements of targets accessible by RV will be beneficial to compare the capabilities of both techniques and to assess their future complementarity.


\paragraph{\textit{MassSpec}'s application domain}

See Main Text (Section \textit{MassSpec}'s Applications).

\paragraph*{\textit{JWST}-\textit{EChO} synergy}

See Main Text (Section Discussion).

\paragraph{No false molecular detections}

\textit{MassSpec}'s applications do not lead to false molecular detections. As emphasized previously, the application of \textit{MassSpec} requires a planet's transmission spectrum of sufficient quality (i.e., SNR and spectral resolution) to identify the major molecular species. Once these criteria are met, the specificity of the species' signature in extinction (shown in panels A of Figs.\,\ref{fig:mini_neptune_JWST}-\ref{fig:nitrogen_world_20m}) prevents from the false positive detection of atmospheric species. As an example, we try to retrieve ozone that was not part of the synthetic atmospheres and it is detected in none of our retrieval. 

\paragraph{On the use of species' mixing ratio in atmospheric retrieval method}

Species' mixing ratios are not adequate parameters for atmospheric retrieval method. The reason is that the uncertainty on mixing ratios encompasses the uncertainty on the number densities and the pressure. We demonstrate that pressure and number densities are the atmospheric parameters embedded in a planet's transmission spectrum. This implies that pressure and mixing ratios are correlated in the context of transmission spectroscopy (Fig.\,\ref{fig:pressuremix}.A). Hence, the posterior probability distribution of mixing ratios are the results of the convolution of the number densities' and pressure's PPDs (Fig.\,\ref{fig:pressuremix}.B), i.e., the uncertainty on the mixing ratios combine the uncertainty on the number densities and the pressure. Therefore, the significance of a molecular detection based on the mixing ratio will be less significant than if based on the number densities, in particular because number densities are more constraint than the pressure.

\clearpage

\section{Dependency of transmission spectroscopy and radial velocity on system parameters}
\label{app:scaling_laws}

Here we introduce \textit{MassSpec}'s sensitivity to the planetary system parameters and show \textit{MassSpec}'s complementarity with the radial velocity method to yield planetary mass measurements. For that purpose, we derive scaling laws to reveal the main dependency of each method's signal on the properties of the observed planetary system, such as the planet's semi-major axis. 

\subsection{Intensity of the RV and Transmission Signals}
\label{app:scaling_laws_signals}

 The signal targeted by the RV method, the RV shift ($K_{\star}$), can be expressed as
\begin{eqnarray}
	K_{\star} = M_p\sqrt{\frac{G}{(M_p+M_{\star})a}}\frac{\sin i}{\sqrt{1-e}},
	\label{eq:RV_shift}
\end{eqnarray}
where $M_{\star}$, $M_p$, $a$, $i$, and $e$ are the host star's mass and the planet's mass, orbital semi-major axis, inclination, and eccentricity \cite{Murray2010}. (We do not discuss here the effects of $i$ and $e$.)
 On the other hand, the signal in transmission depends solely on the area of the opaque atmospheric annulus, $2\pi R_{p,0}h_{eff}(\lambda)$, and the host star spectral radiance $B_{\lambda}(T_{\star})$ (where $B_{\lambda}$ is the Planck function and $T_{\star}$ the star's effective temperature). Using Eq.\,\ref{eq:h_eff_in_text}, we can write that $h_{eff}(\lambda)\propto H$---for active molecular bands $A_\lambda \geq 10^3$ therefore $h_{eff}(\lambda) = nH$ with $n \geq 6$. $H=kT/\mu g$ where T can be approached by the planet's equilibrium temperature at first order, $T_{eq} = T_{\star}(R_{\star}/a)^{0.5}[f'(1-A_B)]^{0.25}$ where $R_{\star}$ is the star's radius and $a$, $f'$, and $A_B$ are respectively the semi-major axis, a parameter for the heat redistribution in the planet's atmosphere ($f'=1/4$ if the heat deposited by the stellar radiation is uniformly distributed, $f'=2/3$ for a tidally-locked planet without atmospheric advection) and the Bond albedo \cite{Seager2010}. By rewritting the planet's surface gravity as $g = 4\pi G \rho_p R_p/3$ ($\rho_p$ and $R_p$ are the planet's density and radius), we can summarize how the system parameters affect the transmission spectrum and the radial velocity signals (Table\,\ref{tab:Tr_RV_app}).

\begin{table}[!h]
\caption{Dependency of transmission spectroscopy and radial velocity to system parameters. \label{tab:Tr_RV_app}}
	\centering
	\setlength{\extrarowheight}{-0pt}
	\setlength{\tabcolsep}{2pt}

	\begin{tabular}{c|c|c}
	
	\hline\hline
	\textbf{Signal}& \textbf{Planetary parameters} & \textbf{Stellar parameters}\\
	\hline
Transmission spectrum & $\rho_p^{-1} \mu^{-1} a^{-0.5}$ & $B_{\lambda}(T_{\star})T_{\star}R_{\star}^{0.5}$\\
RV shift & $M_p a^{-0.5}$ & $M_{\star}^{-0.5}$\\
	\hline

	\end{tabular}
	
\end{table}

Table\,\ref{tab:Tr_RV_app} highlights that signals in transmission are more intense for low-density planets and atmospheres and bright or large stars, while the radial velocity method is ideal for massive planets and low-mass stars. In particular, bright and large stars increase the atmospheric temperature, hence the atmospheric scale height, for a fixed planetary density, atmospheric composition, and semi-major axis. 

\subsection{Sensitivity of a Transmission Spectrum SNR}
\label{app:scaling_laws_snr}

In order to derive the actual scaling law of transmission spectrum SNR---and the sensitivity of \textit{MassSpec}---one have to account for the observational parameters. The overall significance of a signal in transmission and \textit{MassSpec} capabilities scale as
\begin{eqnarray}
	SNR_{ST} \propto \frac{\frac{2\pi R_{p,0}h_{eff}(\lambda)}{d^2} t A \eta B_{\lambda}(T_{\star})}{\sqrt{\frac{\pi R_{\star}^2}{d^2}t A \eta B_{\lambda}(T_{\star})} },
	\label{eq:SNRt_general_scalinglaw_0}
\end{eqnarray}
where the numerator and the denominator relate, respectively, to the number of photons blocked by the planet atmosphere and the uncertainty on the baseline---which corresponds to the square root of the total number of photon emitted by the host star and collected out-of-transit in the same spectral band (Poisson process), respectively. $t$ is the observation time, $A$ and $\eta$ are the telescope's collecting area and total optical throughput, and $d$ is the host star's distance to Earth. Eq.\,\ref{eq:SNRt_general_scalinglaw_0} can be rewritten as
\begin{eqnarray}
	SNR_{ST} \propto T_{\star} [t A \eta B_{\lambda}(T_{\star})]^{0.5} (R_{\star}^{0.5}d \rho_p \mu a^{0.5})^{-1}.
	\label{eq:SNRt_general_scalinglaw}
\end{eqnarray}
For a given planet ($R_{p,0}$,$M_{p}$,$\mu$,$a$, and $T$ fixed), Eq.\,\ref{eq:SNRt_general_scalinglaw} can be rewritten as
\begin{eqnarray}
	SNR_{ST} \propto \frac{\sqrt{t A \eta B_{\lambda}(T_{\star})}}{R_{\star} d}.
	\label{eq:SNRt_general_scalinglaw_fixed_planet}
\end{eqnarray}

Eq.\,\ref{eq:SNRt_general_scalinglaw_fixed_planet} highlights the interest of M dwarfs for obtaining high-SNR transmission spectrum (or transit light-curve) for specific planet properties (e.g., $T_{eq} \sim 300$ K when searching for habitable planets). Eq.\,\ref{eq:SNRt_general_scalinglaw_fixed_planet} shows that for a given planet, the signal significance scales as $\sqrt{B_{\lambda}(T_{\star})}/(R_{\star} d)$. Fig.\,\ref{fig:key_ratio_fixed_planet} shows the ratio $\sqrt{B_{\lambda}(T_{\star})}/R_{\star}$---normalized for a Sun-like star---as a function of the stellar effective temperature. For stars with earlier spectral types than M2V, the significance is independent of the host-star type. However, the significance increase substantially for M dwarfs. We take advantage of this favorable properties of late M dwarfs to show that \textit{MassSpec} will enable to fully characterize Earth-sized planets using \textit{JWST}'s and/or \textit{EChO}'s observations of Earth-sized planets (Section \,\ref{app:results}).

\clearpage

\section{Introduction to rational functions}
\label{app:TF}

A rational function, $f(.)$, is a function that can be expressed as the ratio of polynomials, $P(.)/Q(.)$. The zeros and poles of $f(.)$ are the zeros of $P(.)$ and the zeros of $Q(.)$, respectively. Rational functions are used in various fields of science and engineering (signal processing, acoustics, aerodynamics, structural dynamics, electronic circuitry, control theory, etc.) as mathematical representation of the relation of the inputs ($i(.)$) and outputs ($o(.)$) of a system, called the transfer function of the system. The transfer function of a system ($H(s)$) is the linear mapping of the Laplace/Fourier transform of its inputs ($I(s) = \mathcal{L}(i(.))$) to the Laplace/Fourier transform of its outputs ($O(s) = \mathcal{L}(o(.))$), i.e., $O(s) = H(s)I(s)$ where $s$ is a spatial or temporal frequency if $i$ and $o$ are functions of space or time, respectively. As an example, the transfer function of imaging devices is the Fourier transform of the point spread function (PSF)---because the theoretical input leading to the PSF is a spatial impulse (i.e., a Dirac function) in the field of view and the Fourier transform of an impulse is one, therefore $\mathcal{L}(PSF) = H(s)$. 

The transfer function of a system relates directly to the differential equations used to represent mathematically the system. Therefore, the poles and zeros of a transfer function characterize the behavior of a system. For example, the differential equation of a second order system is
\begin{equation}
\ddot{x}+2 \zeta \omega_n \dot{x}+ \omega_n^2 x=0,
\end{equation}
where $\zeta$ and $\omega_n$ are the damping ratio and the natural frequency of the system, respectively---e.g., for a mass-spring-dashpot system $\zeta = c/(1\sqrt{km})$ and $\omega_n = \sqrt{k/m}$ where $m$, $c$, and $k$ are the mass, the spring constant, and the damping coefficient, respectively. The transfer function of a second order system is 
\begin{equation}
H(s) = K \frac{\omega_n^2}{s^2+2 \zeta \omega_n s+ \omega_n^2} ,
\end{equation}
where $K$ is the system gain (K = 1 for the mass-spring-dashpot system) and $s = j\omega$ where $j$ is the imaginary unit (see Fig.\,\ref{fig:rational_function} for more details). Therefore, at frequencies low compared to $\omega_n$, $H(j\omega) \approx K$, meaning that the response of a mass-spring-dashpot system to a low-frequency input, $i(t) = A\exp(j\omega t)$, is the input, i.e. $o(t) = i(t)$. At high frequencies, $H(j\omega) \approx K/(j\omega)^2 = -K/\omega^2$ meaning that the response is in opposition of phase with the input and its amplitude is proportional to $\omega^{-2}$. 

The position of a transfer function poles and zeros in the complex field ($\mathbb{C}$) define the domains of different dependency regimes for the transfer function (Fig.\,\ref{fig:rational_function}). In the neighborhood of a zero, the exponent of the transfer function dependency to its variable increases by 1, while it decreases by 1 in the neighborhood of a pole. Hence, the overall shape of transfer function relates directly to its formulation. Therefore, from simulations or measurements, it is possible to derive the appropriate formulation to represent mathematically a system using a rational function.

\clearpage

\begin{figure}[!p]
 \centering
  \begin{center}
    \includegraphics[trim = 00mm 00mm 00mm 10mm,clip,width=15cm,height=!]{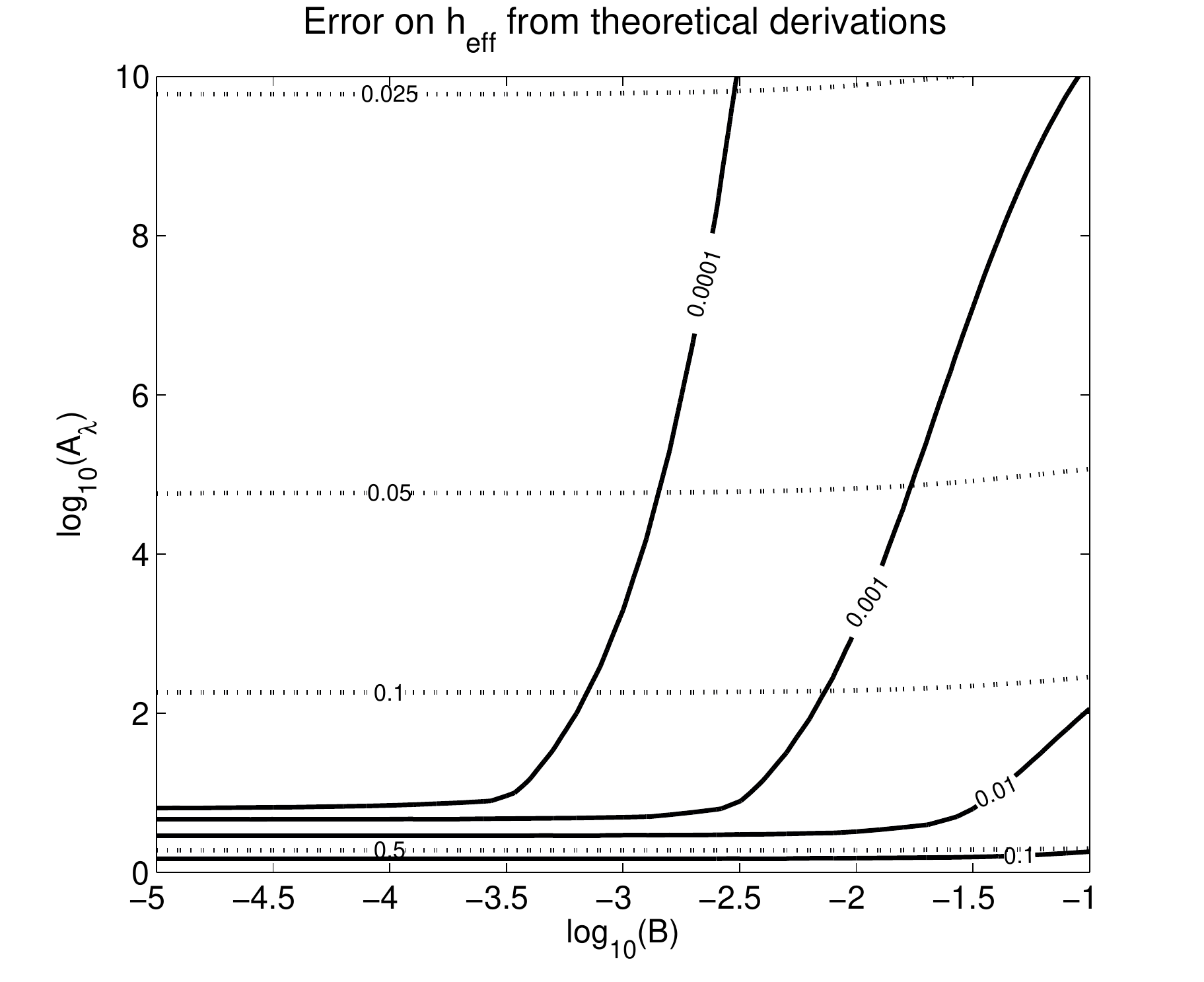}
  \end{center}
  \vspace{-0.7cm}
  \caption{Relative errors on the apparent atmosphere height ($h_{eff}(\lambda)$) using the analytic solution to Eq.\,\ref{eq:transmission_spectrum_h} (or Eq.\,\ref{eq:def_yeff}) for extinction processes with a cross-section independent of $T$ and $p$ (such as Rayleigh scattering). Relative errors for the zeroth-order derivation $[h_{eff}(\lambda) = z(\tau(\lambda)=1)]$ correspond to the dotted lines, and the ones for our derivation $h_{eff}(\lambda) = [z:\tau(z,\lambda) = e^{-\gamma_{EM}}]$ are the thick lines.}
  \vspace{-0.0cm}
  \label{fig:Error_on_cst_cross_sect_derivation}
\end{figure}

\begin{figure}[!p]
 \centering
  \begin{center}
    {\hspace{-3cm}\includegraphics[trim = 15mm 00mm 20mm 0mm,clip,width=19cm,height=!]{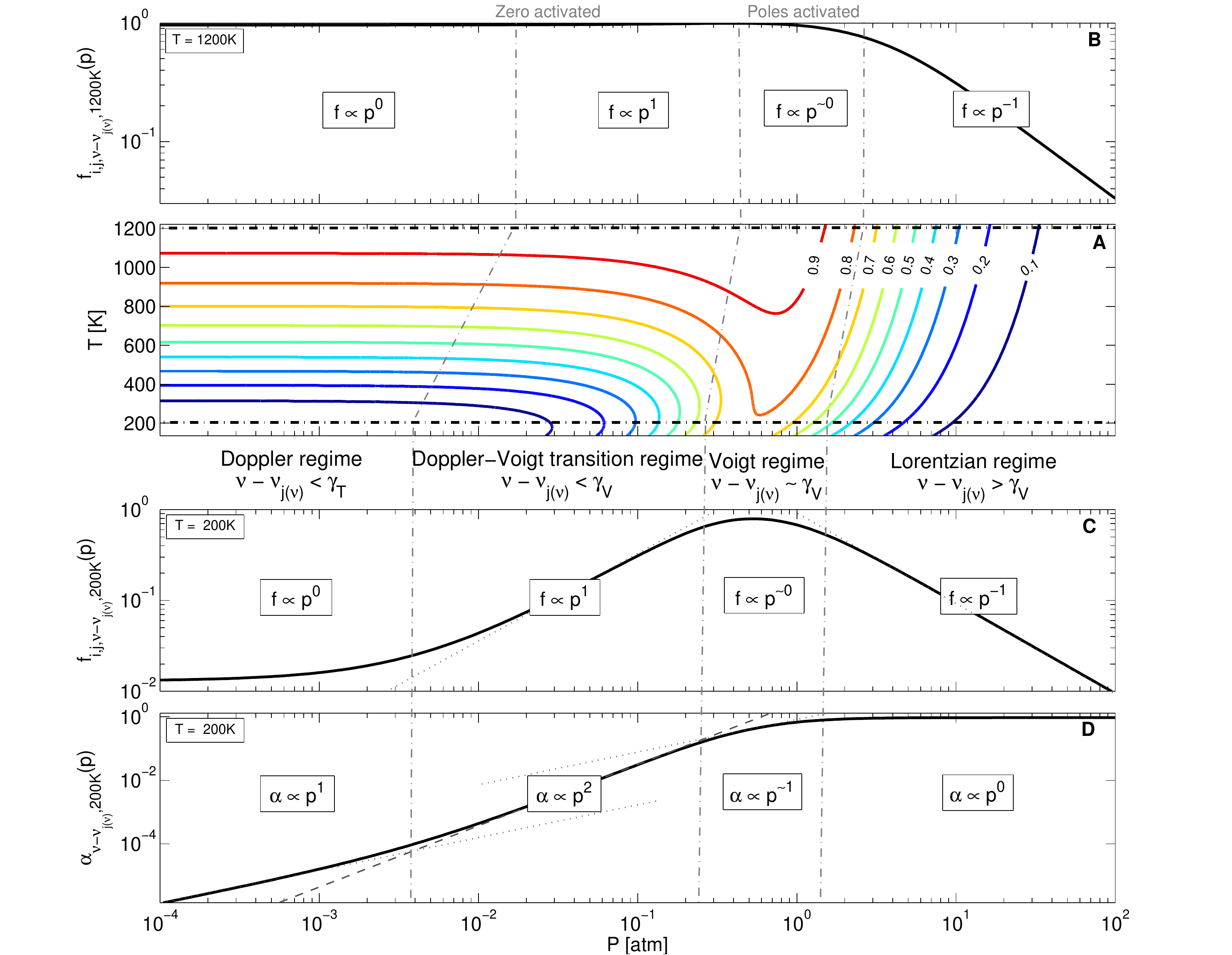}\hspace{-0cm}}
  \end{center}
  \vspace{-0.7cm}
  \caption{A line profile ($f_\nu$) depends on the pressure ($p$) as a rational function at fixed temperature ($T$) and frequency ($\nu$). (\textbf{A}) Dependency of $f_\nu$ on $T$ and $p$, at fixed $\nu$, that shows the four domains of different dependency regimes of $f_\nu$ on $p$ whose boundaries are $T$-dependent (gray dot-dash lines). The black dot-dash lines represents the position of the slices in the $\left\lbrace T-p-f\right\rbrace$ space used to highlight that $f_\nu$ behaves as a rational function of $p$, at $\{T,\nu\}$ fixed. Planels \textbf{B} and \textbf{C} present the slices at $T=1200$K and $T=200$K, respectively. These slices show that $f_\nu$ behaves as a rational function of $p$ with a zero and a pair of conjugated zeros (Fig.\,\ref{fig:rational_function}). In particular, the absolute value of the zero is less than the poles', as underscored by the sequential transition from the following dependency regimes $\propto p^{0}$, $\propto p^{1}$, $\propto p^{\sim0}$, and $\propto p^{-1}$ with increasing $p$---the dotted and the dashed lines represent a slope of 1 and 2, respectively. (\textbf{D}) Dependency of the absorption coefficient ($\alpha_\nu$) on $p$, at $T=200$K. The exponent of the $\alpha_\nu$ dependency on $p$ increases by one compared to $f_\nu$ dependency on $p$. The exponent increase by one because of $\alpha_\nu$'s additional zero at $p=0$ that originates from the number density.}
  \vspace{-0.0cm}
  \label{fig:T_p_dependence_of_line_profile}
\end{figure}

\begin{figure}[!p]
 \centering
  \begin{center}
    \includegraphics[trim = 00mm 00mm 00mm 00mm,clip,width=15cm,height=!]{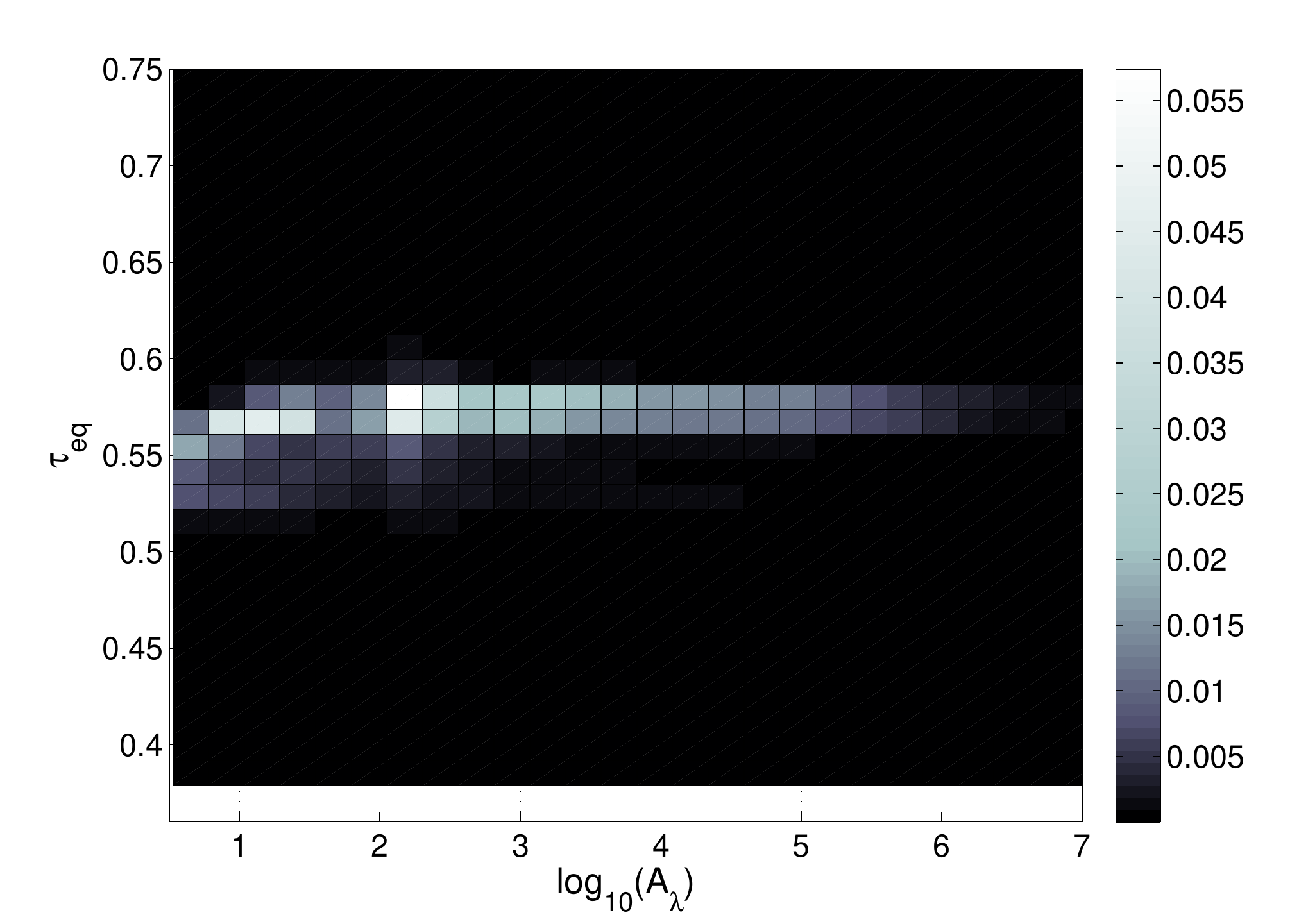}
  \end{center}
  \vspace{-0.7cm}
  \caption{Distribution of $\tau_{eq}(\lambda)$ as a function of $\log_{10}(A_{\lambda})$ for an isothermal-isocomposition Earth. $\sim$99$\%$ of the active bins have a $\tau_{eq}(\lambda)\approx \exp^{-\gamma_{EM}}$.}
  \vspace{-0.0cm}
  \label{fig:tau_distribution_isothermal_and_real_Earth}
\end{figure}

\begin{figure}[!p]
 \centering
  \begin{center}
    \includegraphics[trim = 00mm 00mm 00mm 00mm,clip,width=15cm,height=!]{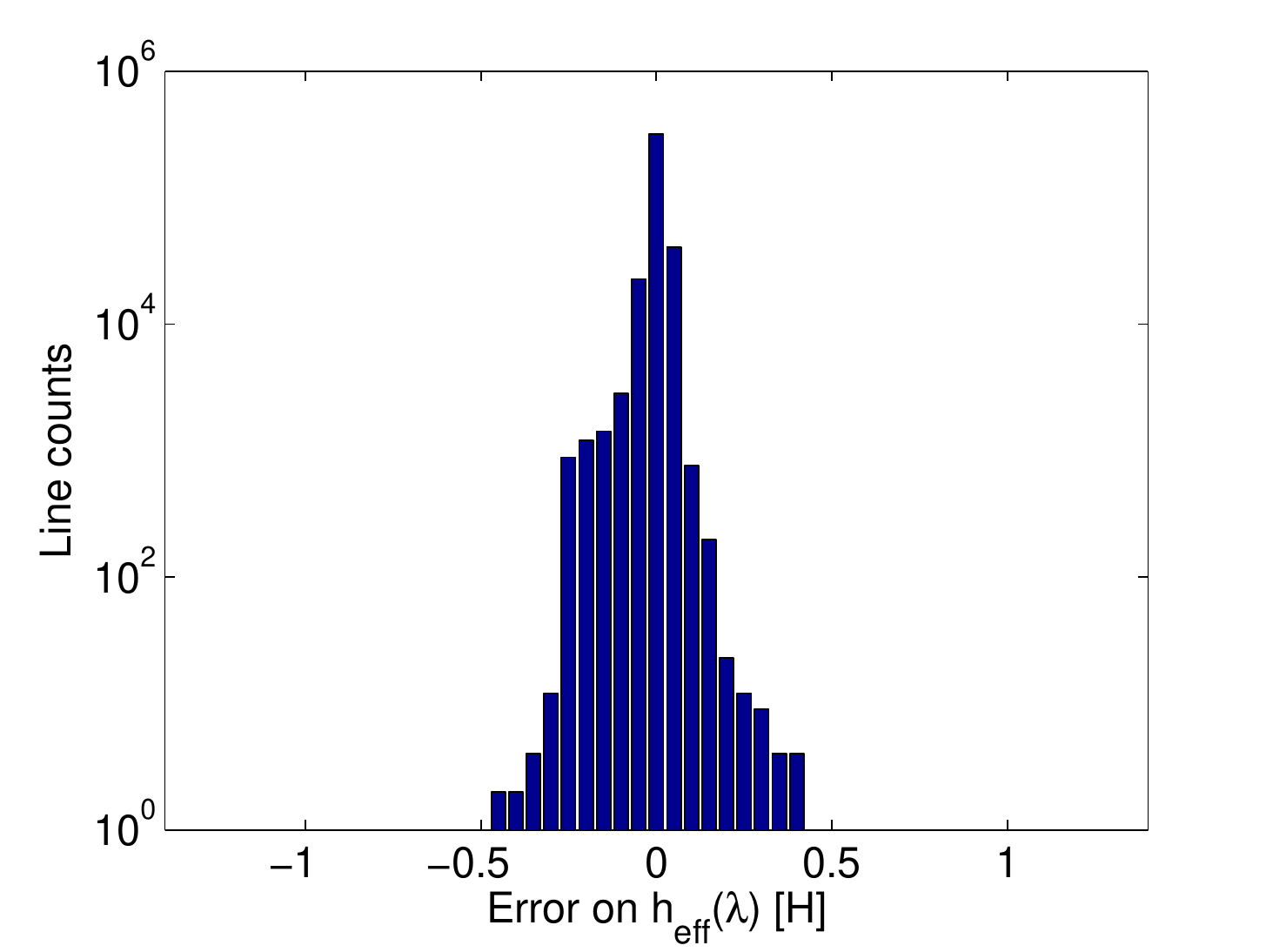}
  \end{center}
  \vspace{-0.7cm}
  \caption{Distribution of the error on $h_{eff}(\lambda)$ (expressed in scale heights) when estimated from $h_{eff}(\lambda) = [z:\tau(z,\lambda) = e^{-\gamma_{EM}}]$ for an isothermal-isocomposition Earth. The error is below $3\%$ for 99.7$\%$ of the lines (3$\sigma$ confidence interval). If the error is below 3$\%$, the error on the simulated apparent height is below 250 meters. This highlights how well transmission spectrum can be modeled by solving $h_{eff}(\lambda) = [z:\tau(z,\lambda) = e^{-\gamma_{EM}}]$.}
  \vspace{-0.0cm}
  \label{fig:error_on_h_eff_iso_Earth}
\end{figure}

\begin{figure}[!p]
 \centering
  \begin{center}
    \includegraphics[trim = 00mm 00mm 00mm 00mm,clip,width=15cm,height=!]{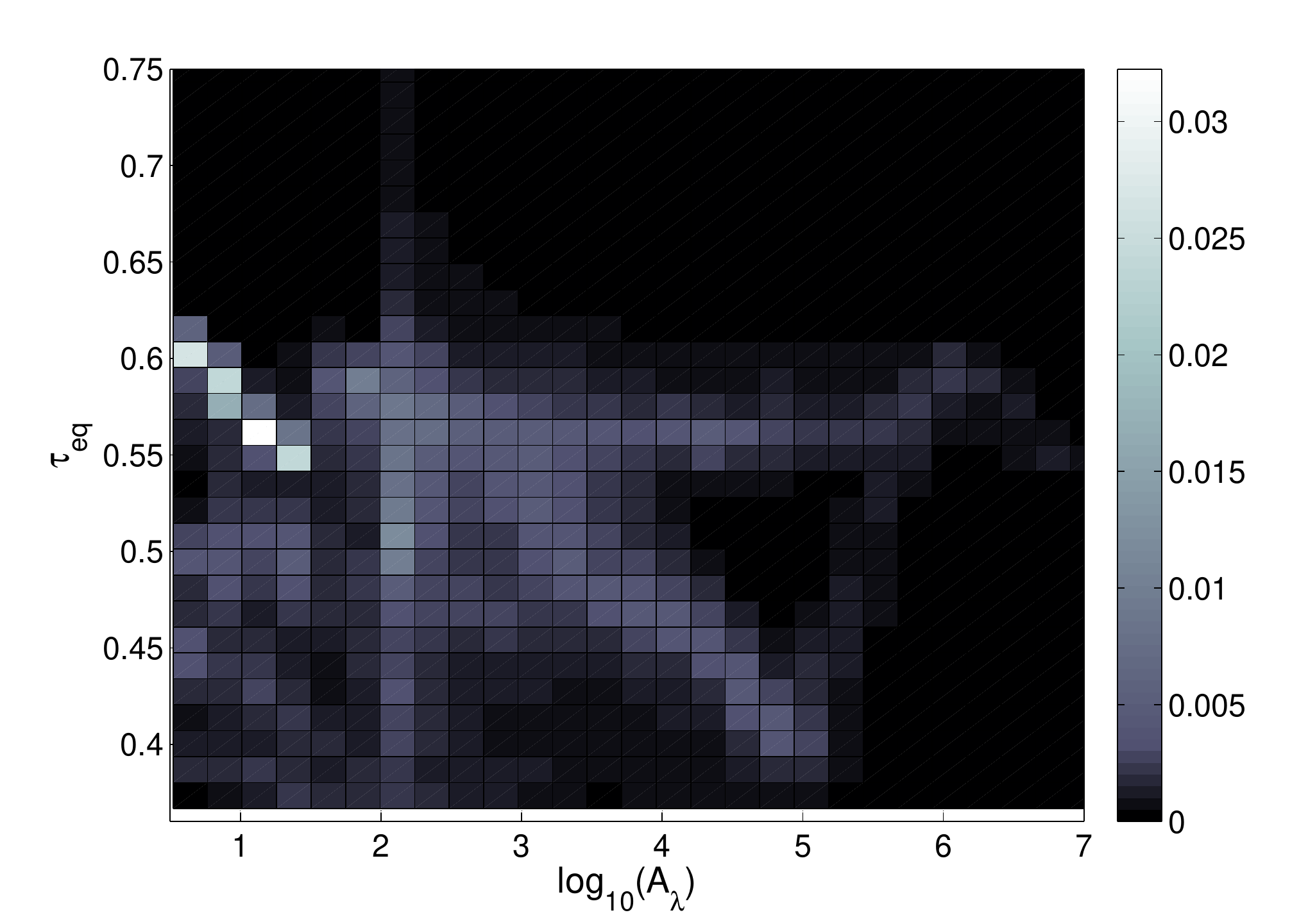}
  \end{center}
  \vspace{-0.7cm}
  \caption{Distribution of $\tau_{eq}(\lambda)$ as a function of $\log_{10}(A_{\lambda})$ for Earth---temperature-pressure-mixing ratio profiles from \cite{Cox2000}. $\sim$50$\%$ of the active bins have a $\tau_{eq}(\lambda)\approx \exp^{-\gamma_{EM}}$.}
  \vspace{-0.0cm}
  \label{fig:tau_eq_vs_a_lambda_Earth}
\end{figure}

\begin{figure}[!p]
 \centering
  \begin{center}
    \includegraphics[trim = 00mm 00mm 00mm 00mm,clip,width=15cm,height=!]{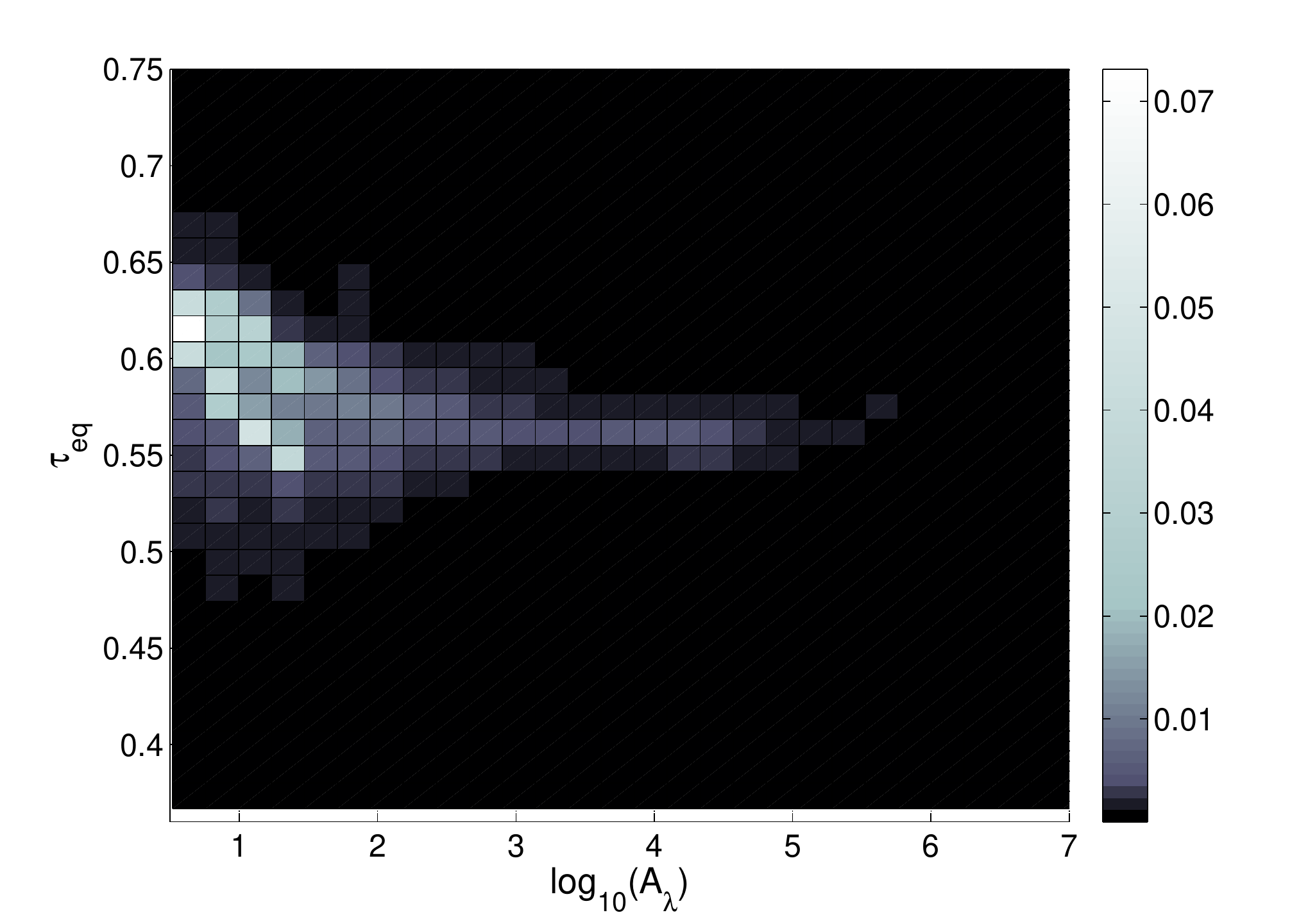}
  \end{center}
  \vspace{-0.7cm}
  \caption{Distribution of $\tau_{eq}(\lambda)$ as a function of $\log_{10}(A_{\lambda})$ for Earth with no water in its atmosphere---temperature-pressure-mixing ratio profiles from \cite{Cox2000}. $\sim$80$\%$ of the active bins have a $\tau_{eq}(\lambda)\approx \exp^{-\gamma_{EM}}$.}
  \vspace{-0.0cm}
  \label{fig:tau_eq_vs_a_lambda_Earth_without_water}
\end{figure}

\begin{figure}[!p]
 \centering
  \begin{center}
    \includegraphics[trim = 00mm 00mm 00mm 00mm,clip,width=15cm,height=!]{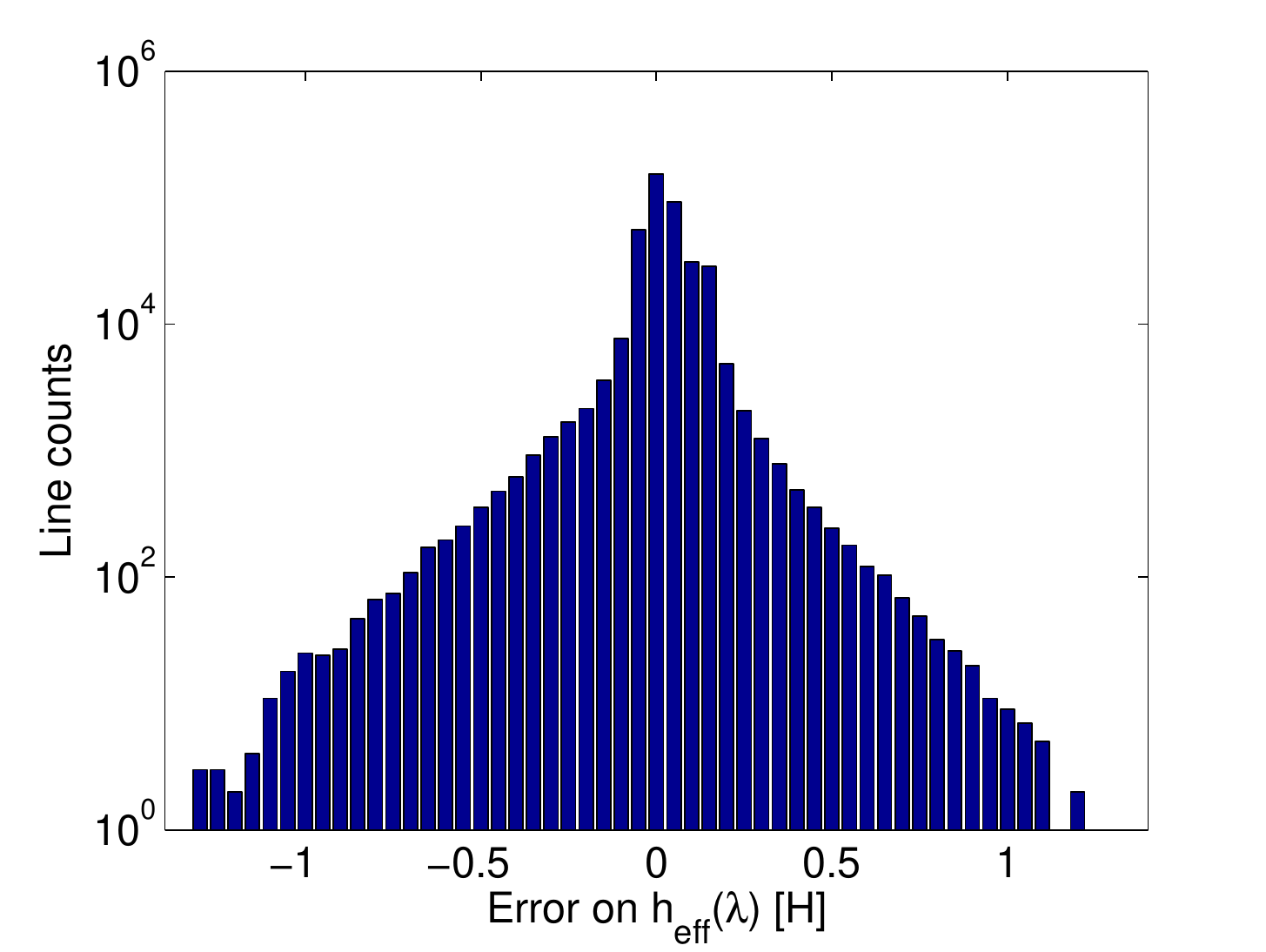}
  \end{center}
  \vspace{-0.7cm}
  \caption{Distribution of the error on $h_{eff}(\lambda)$ (expressed in scale heights) when estimated from $h_{eff}(\lambda) = [z:\tau(z,\lambda) = e^{-\gamma_{EM}}]$ for Earth---temperature-pressure-mixing ratio profiles from \cite{Cox2000}. The error is below $18\%$ for 99.7$\%$ of the lines (3$\sigma$ confidence interval). If the error is below 18$\%$, the error on the simulated apparent height is below 1500 meters. The error is larger than for an isothermal-isocomposition Earth (Fig.\,\ref{fig:error_on_h_eff_iso_Earth}) mainly because water's mixing ratio drops significantly in the troposphere (as highlight by the comparison of Figs\,\ref{fig:tau_eq_vs_a_lambda_Earth} and \ref{fig:tau_eq_vs_a_lambda_Earth_without_water}).}
  \vspace{-0.0cm}
  \label{fig:error_on_h_eff_Earth}
\end{figure}

\clearpage

\begin{figure}[!ht]
  \begin{center}
    \includegraphics[trim = 00mm 00mm 00mm 0mm,clip,width=12cm,height=!]{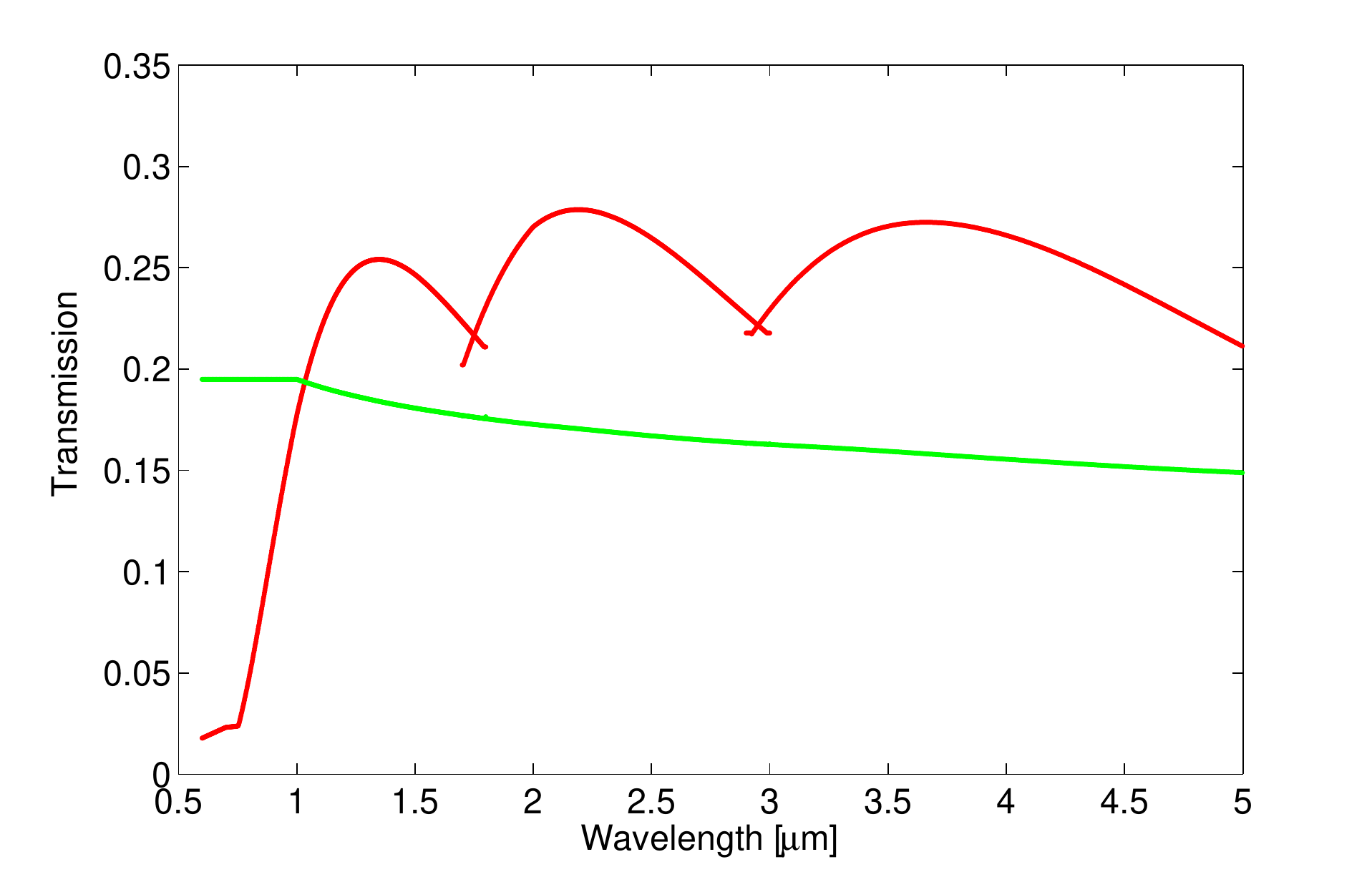}
  \end{center}
  \vspace{-0.6cm}
  \caption{\textit{JWST}/\textit{NIRSpec} optical performance for the medium spectral resolution mode (R = 1000). The red line shows the total optical throughput for each spectral resolution element. The green line shows the flux fraction going to the brightest pixel of each spectral resolution element.}
  \label{fig:nirspec_perfo}
\end{figure}

\begin{figure}[!ht]
  \begin{center}
    \includegraphics[trim = 00mm 00mm 00mm 00mm,clip,width=17cm,height=!]{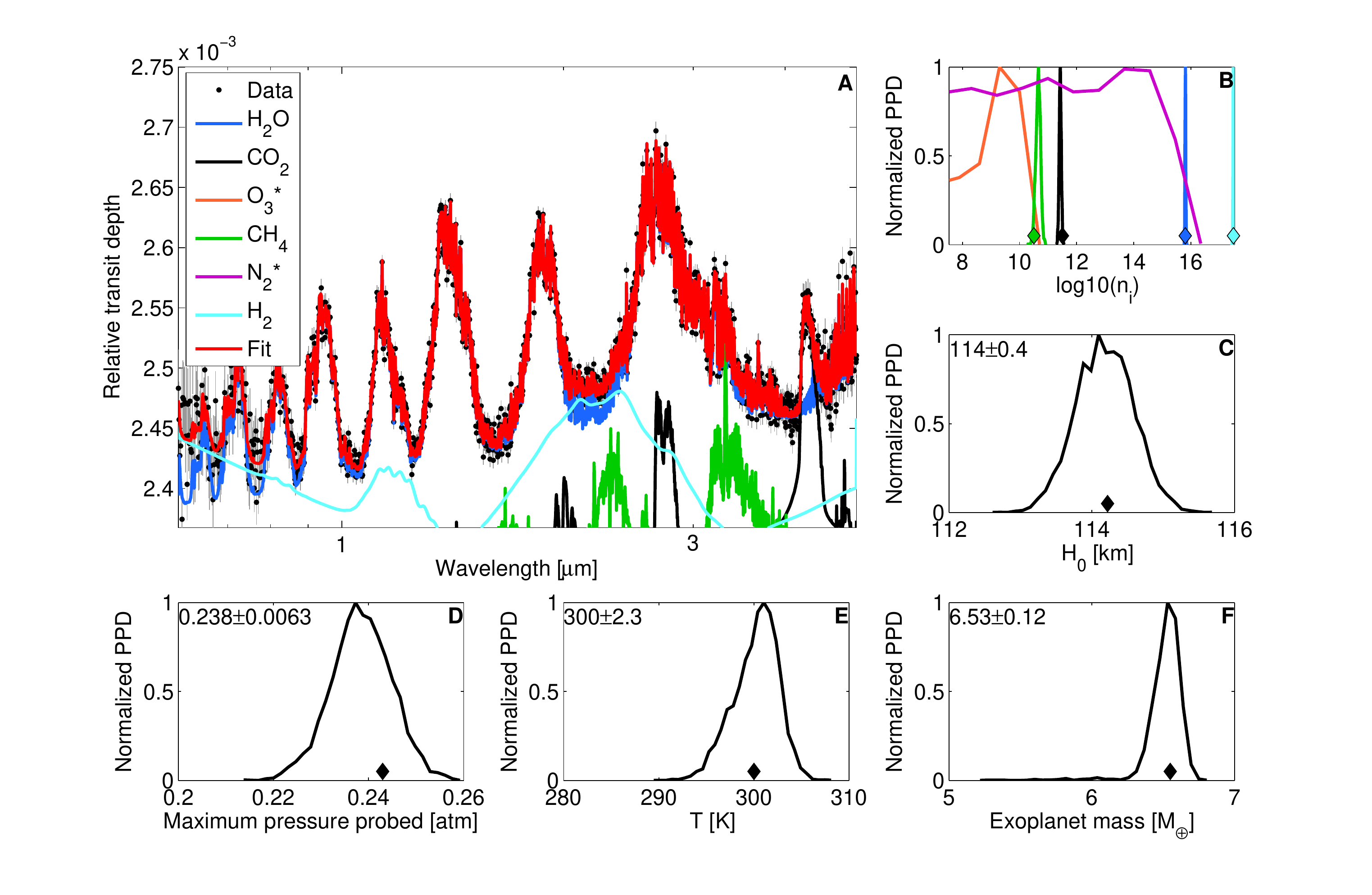}
  \end{center}
  \vspace{-0.6cm}
  \caption{\textit{MassSpec}'s application to the synthetic transmission spectrum of a hydrogen-dominated super-Earth transiting a M1V star at 15 pc as observed with \textit{JWST} for a total of 200 hrs in-transit. The panels show the same quantities as on Fig.\,\ref{fig:MassSpec_results_in_text_ww}. 
The atmospheric properties (number densities, scale height and temperature) are retrieved with high significance yielding to a mass measurement with a relative uncertainty of $\sim2\%$. Note the significant difference between the number density PPD’s of hydrogen, water, carbon dioxide, and methane and those of ozone and nitrogen (\textbf{B}). The latter two gases were not part of the synthetic atmosphere. Ozone and nitrogen are not detected, because no constraints on their mixing ratios can be made. Hydrogen-dominated planets are targets that are particularly favorable for \textit{MassSpec} because their extended atmosphere leads to high-SNR transmission spectra.}
  \label{fig:mini_neptune_JWST}
\end{figure}


\begin{figure}[!ht]
  \begin{center}
    \includegraphics[trim = 00mm 00mm 00mm 00mm,clip,width=17cm,height=!]{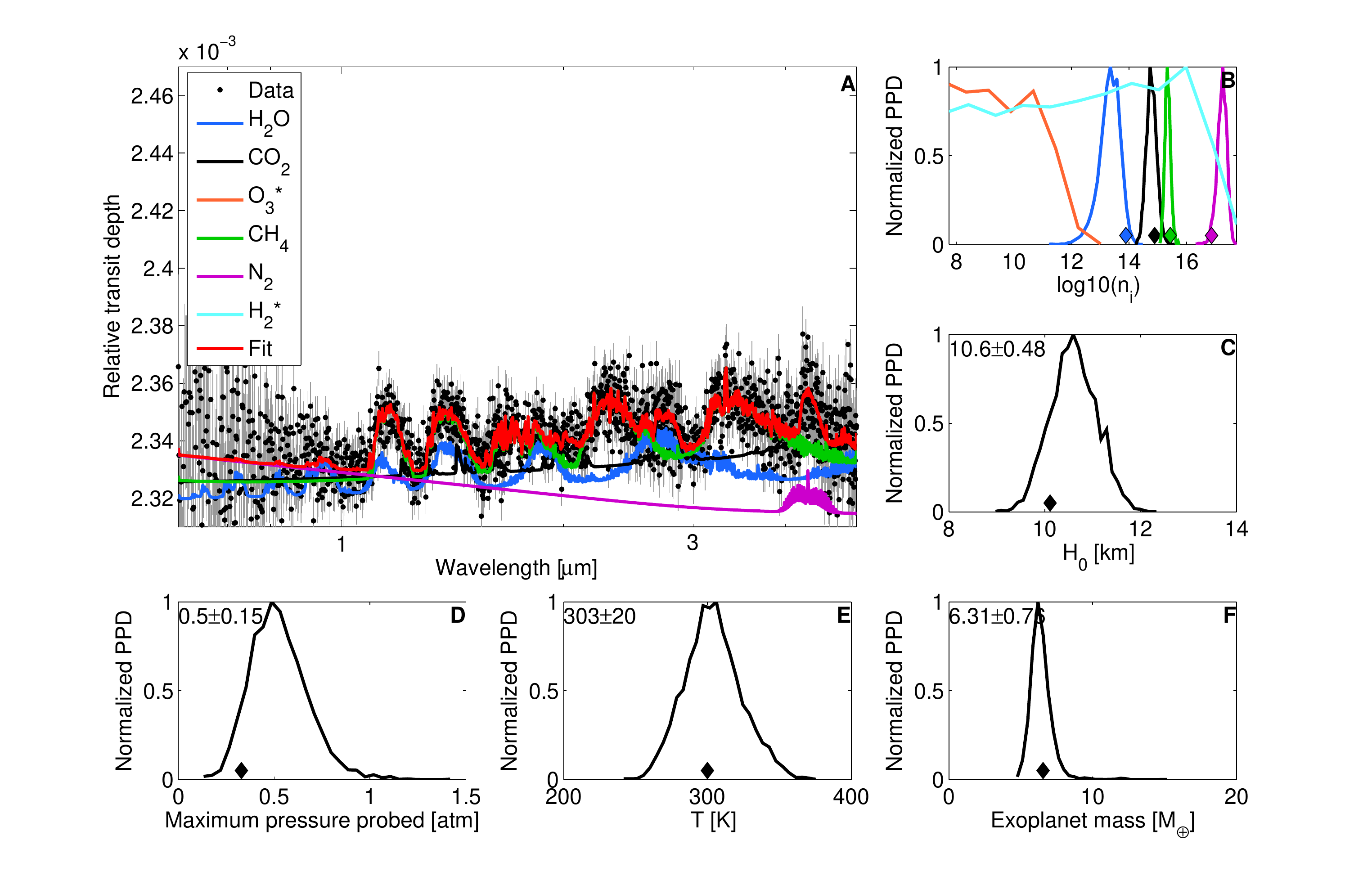}
  \end{center}
  \vspace{-0.6cm}
  \caption{\textit{MassSpec}'s application to the synthetic transmission spectrum of a super-Earth with a nitrogen-dominated atmosphere transiting a M1V star at 15 pc as observed with \textit{JWST} for a total of 200 hrs in-transit. The panels show the same quantities as on Fig.\,\ref{fig:MassSpec_results_in_text_ww}. 
The atmospheric properties are retrieved with high significance yielding to a mass measurement with a relative uncertainty of $\sim15\%$.}
  \label{fig:nitrogen_world_JWST}
\end{figure}

\begin{figure}[!ht]
  \begin{center}
    \includegraphics[trim = 00mm 00mm 00mm 00mm,clip,width=17cm,height=!]{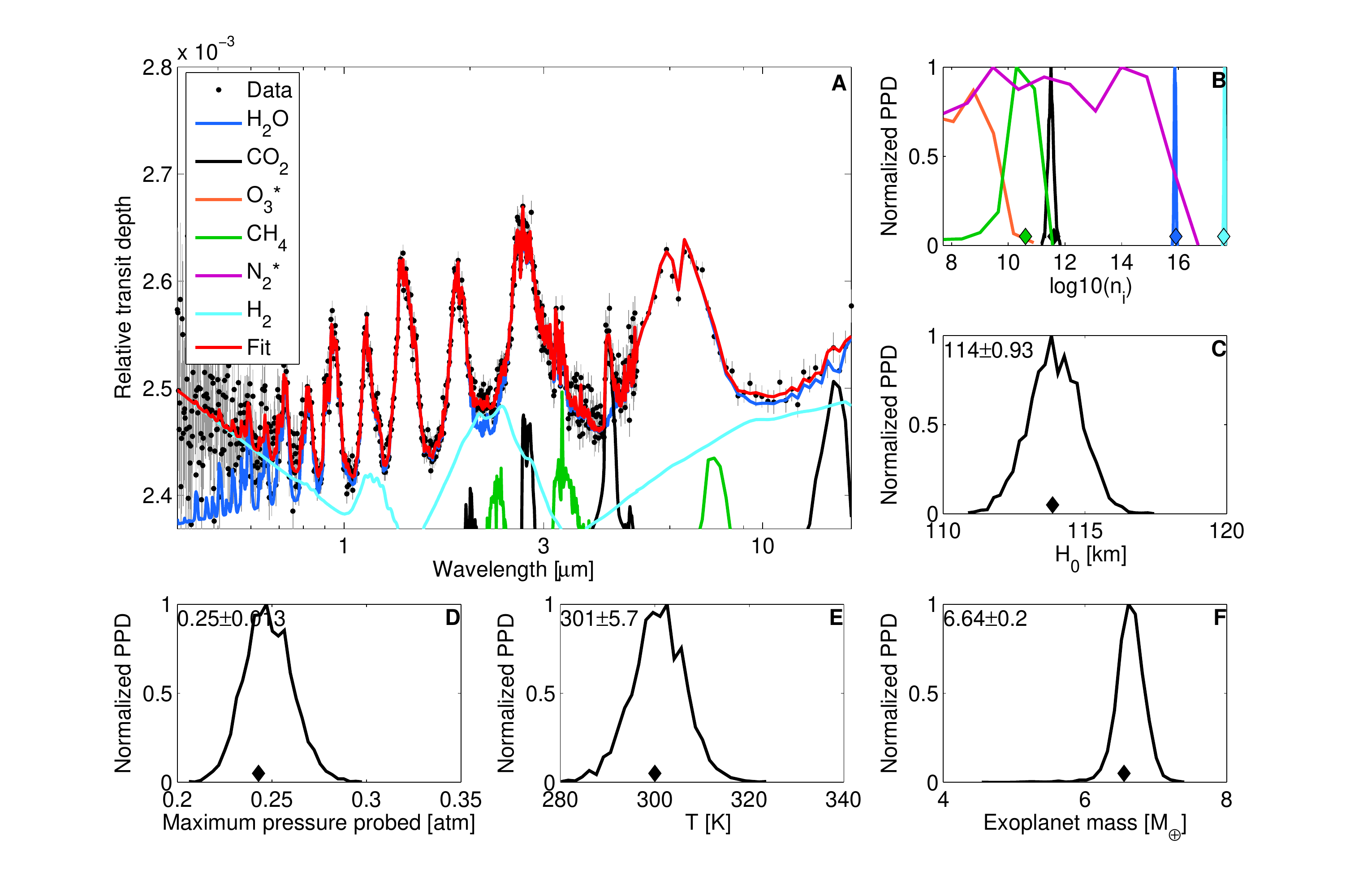}
  \end{center}
  \vspace{-0.6cm}
  \caption{\textit{MassSpec}'s application to the synthetic transmission spectrum of a super-Earth with a hydrogen-dominated atmosphere transiting a M1V star at 15 pc as observed with \textit{EChO} for a total of 200 hrs in-transit. The panels show the same quantities as on Fig.\,\ref{fig:MassSpec_results_in_text_ww}. 
The atmospheric properties are retrieved with high significance yielding to a mass measurement with a relative uncertainty of $\sim3\%$. Note that \textit{EChO}'s capabilities for mass measurements are enhanced by its large spectral coverage that yields the Rayleigh-scattering slope, which is particularly valuable for mass and atmosphere retrieval.}
  \label{fig:mini_neptune_EChO}
\end{figure}

\begin{figure}[!ht]
  \begin{center}
    \includegraphics[trim = 00mm 00mm 00mm 00mm,clip,width=17cm,height=!]{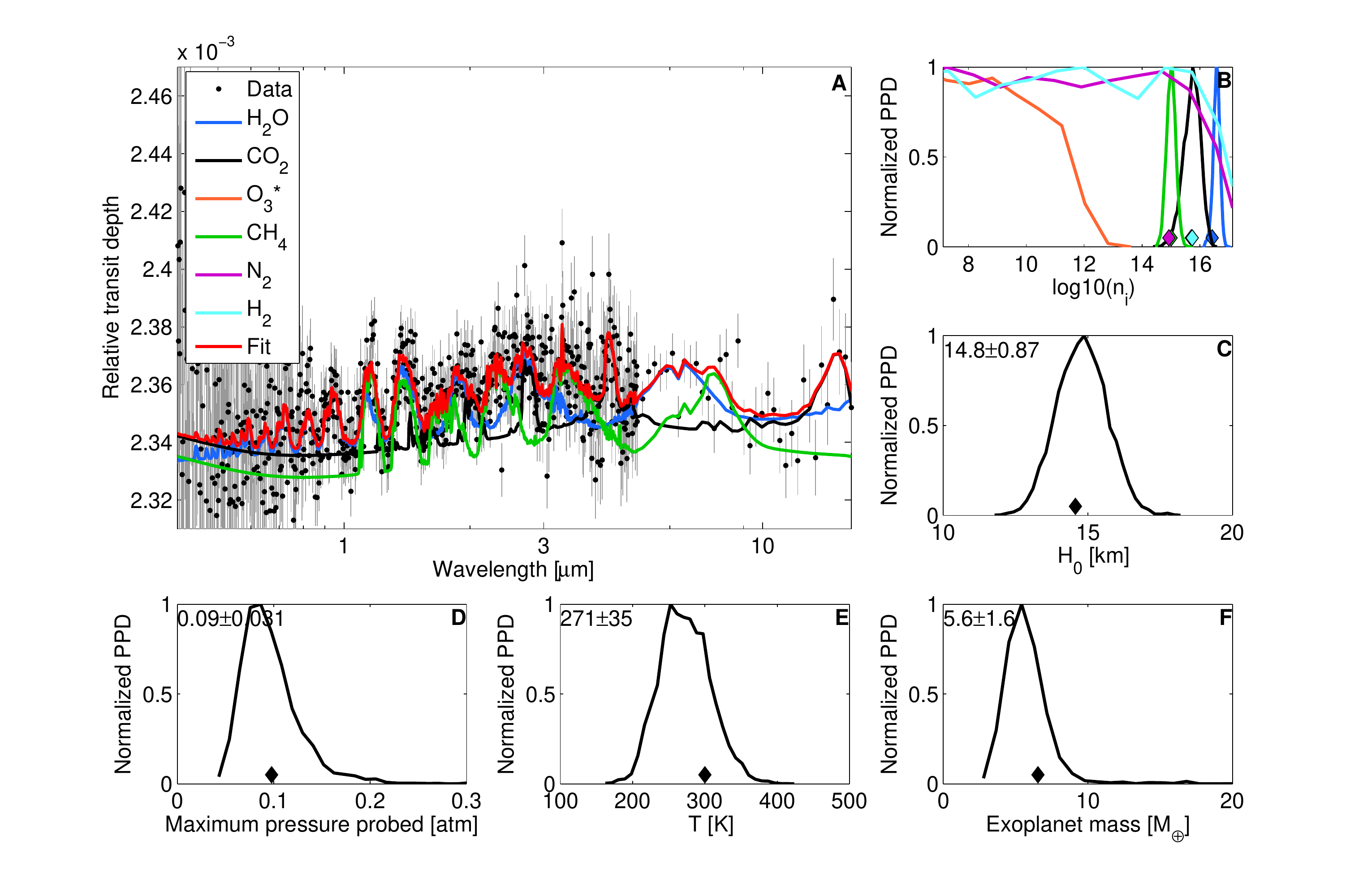}
  \end{center}
  \vspace{-0.6cm}
  \caption{\textit{MassSpec}'s application to the synthetic transmission spectrum of a super-Earth with a water-dominated atmosphere transiting a M1V star at 15 pc as observed with \textit{EChO} for a total of 200 hrs in-transit. The panels show the same quantities as on Fig.\,\ref{fig:MassSpec_results_in_text_ww}.  
The atmospheric properties are retrieved with high significance yielding to a mass measurement with a relative uncertainty of $\sim25\%$.}
  \label{fig:water_world_EChO}
\end{figure}

\begin{figure}[!ht]
  \begin{center}
    \includegraphics[trim = 00mm 00mm 00mm 00mm,clip,width=17cm,height=!]{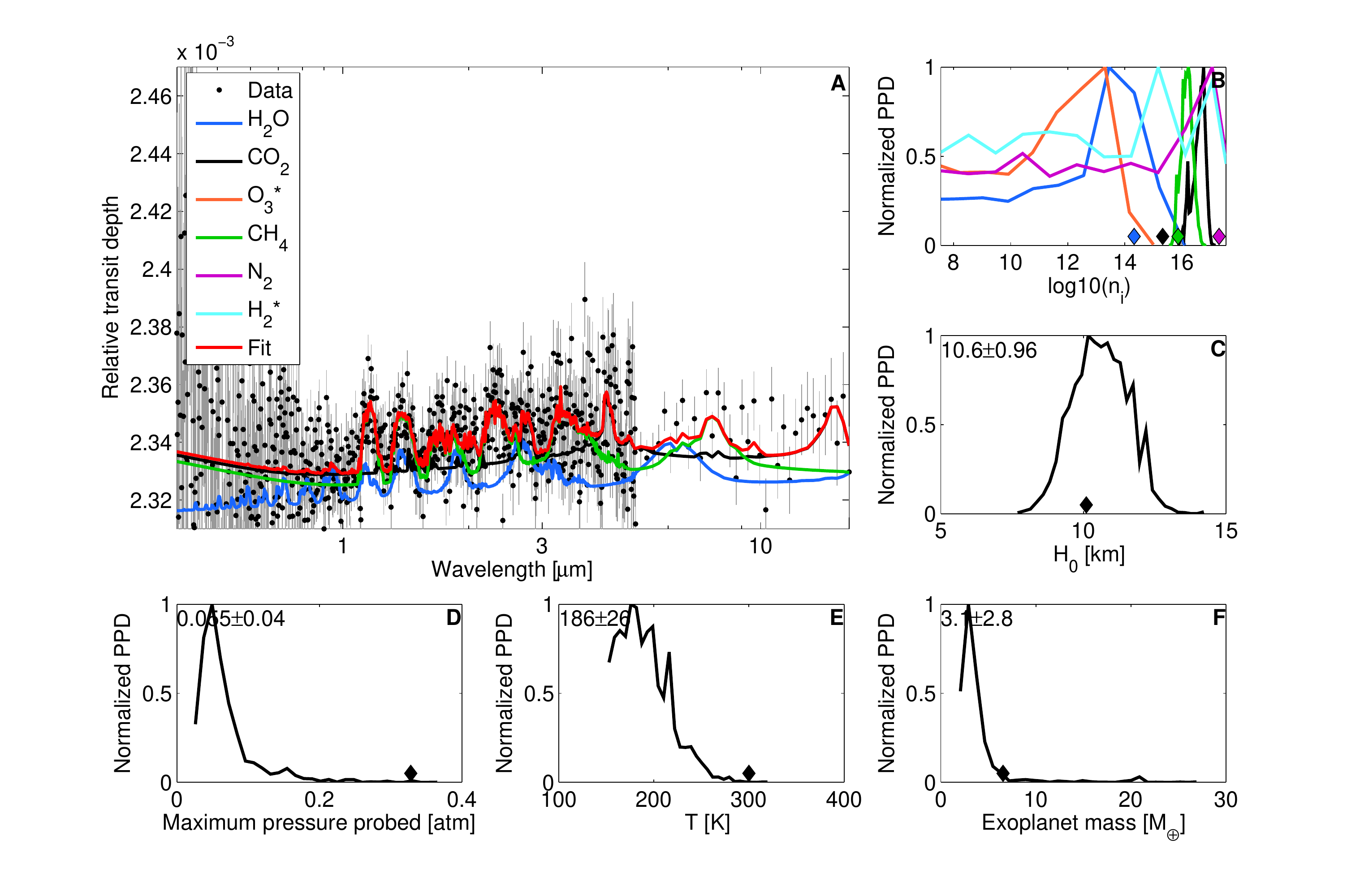}
  \end{center}
  \vspace{-0.6cm}
  \caption{\textit{MassSpec}'s application to the synthetic transmission spectrum of a super-Earth with a nitrogen-dominated atmosphere transiting a M1V star at 15 pc as observed with \textit{EChO} for a total of 200 hrs in-transit. The panels show the same quantities as on Fig.\,\ref{fig:MassSpec_results_in_text_ww}. The planet mass is not retrieved because the data quality does not yield the atmosphere composition and temperature---although the scale height and the signatures of water, methane, and carbon dioxide are retrieved.}
  \label{fig:nitrogen_world_EChO}
\end{figure}

\begin{figure}[!ht]
  \begin{center}
    \includegraphics[trim = 00mm 00mm 00mm 00mm,clip,width=17cm,height=!]{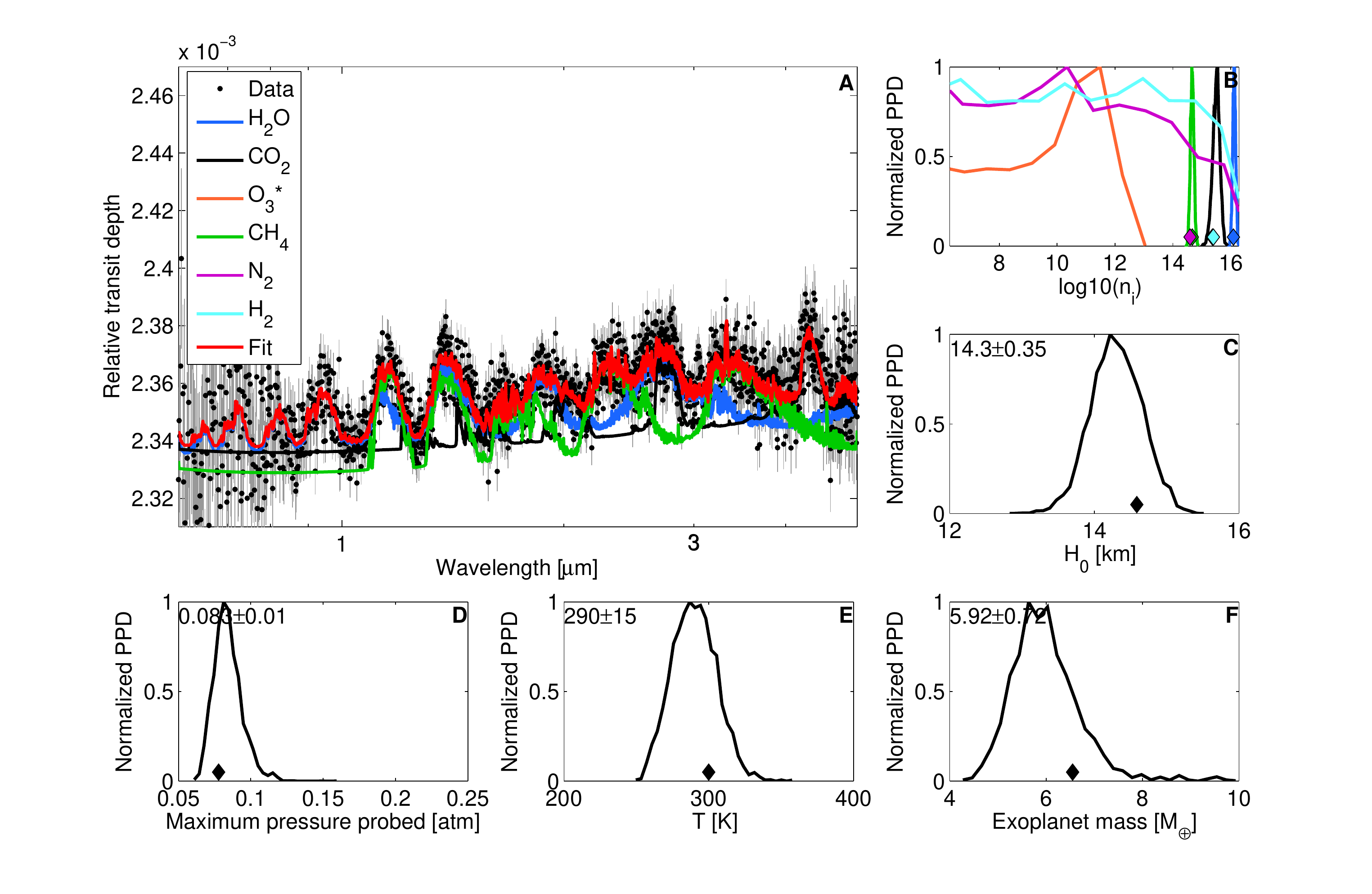}
  \end{center}
  \vspace{-0.6cm}
  \caption{\textit{MassSpec}'s application to the synthetic transmission spectrum of a super-Earth with a water-dominated atmosphere presenting a cloud deck at 100 mbar (such as Venus) transiting a M1V star at 15 pc as observed with \textit{JWST} for a total of 200 hrs in-transit. The panels show the same quantities as on Fig.\,\ref{fig:MassSpec_results_in_text_ww}.
The atmospheric properties are retrieved with high significance yielding to a mass measurement with a relative uncertainty of $\sim10\%$. \textit{MassSpec}'s capabilities are not affected by the 100-mbar cloud deck because transmission spectroscopy does not probe deeper such planet atmosphere (Fig.\,\ref{fig:MassSpec_results_in_text_ww}, panel D).}
  \label{fig:water_world_JWST_C1}
\end{figure}

\begin{figure}[!ht]
  \begin{center}
    \includegraphics[trim = 00mm 00mm 00mm 00mm,clip,width=17cm,height=!]{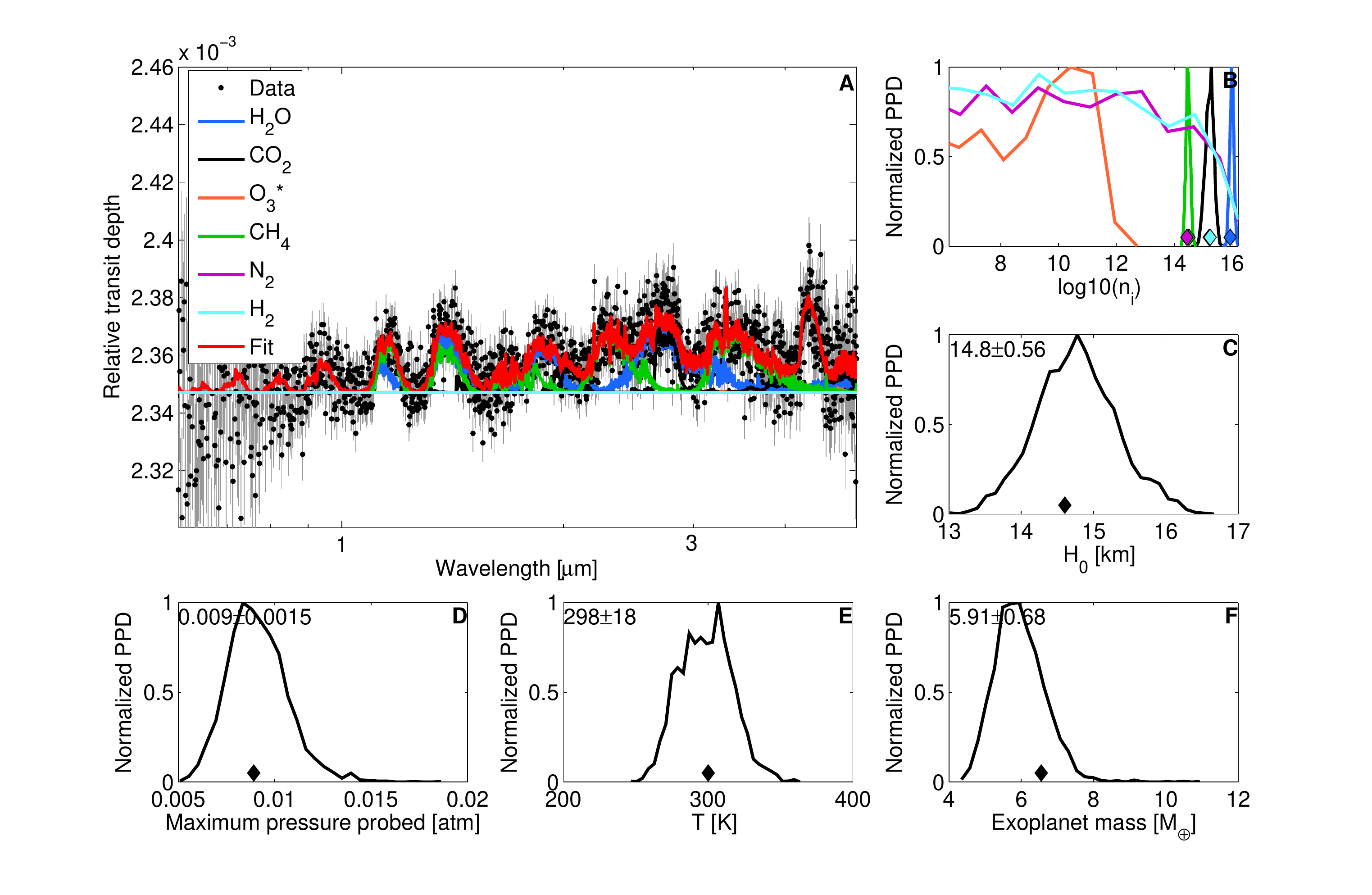}
  \end{center}
  \vspace{-0.6cm}
  \caption{\textit{MassSpec}'s application to the synthetic transmission spectrum of a super-Earth with a water-dominated atmosphere presenting a cloud deck at 10 mbar transiting a M1V star at 15 pc as observed with \textit{JWST} for a total of 200 hrs in-transit. The panels show the same quantities as on Fig.\,\ref{fig:MassSpec_results_in_text_ww}. The atmospheric properties are retrieved with high significance yielding to a mass measurement with a relative uncertainty of $\sim10\%$. \textit{MassSpec}'s capabilities are marginally affected by the 100-mbar cloud deck because a limited fraction of the spectral bins probe deeper than 0.01 atm such planet atmosphere.}
  \label{fig:water_world_JWST_C2}
\end{figure}

\begin{figure}[!ht]
  \begin{center}
    \includegraphics[trim = 00mm 00mm 00mm 00mm,clip,width=17cm,height=!]{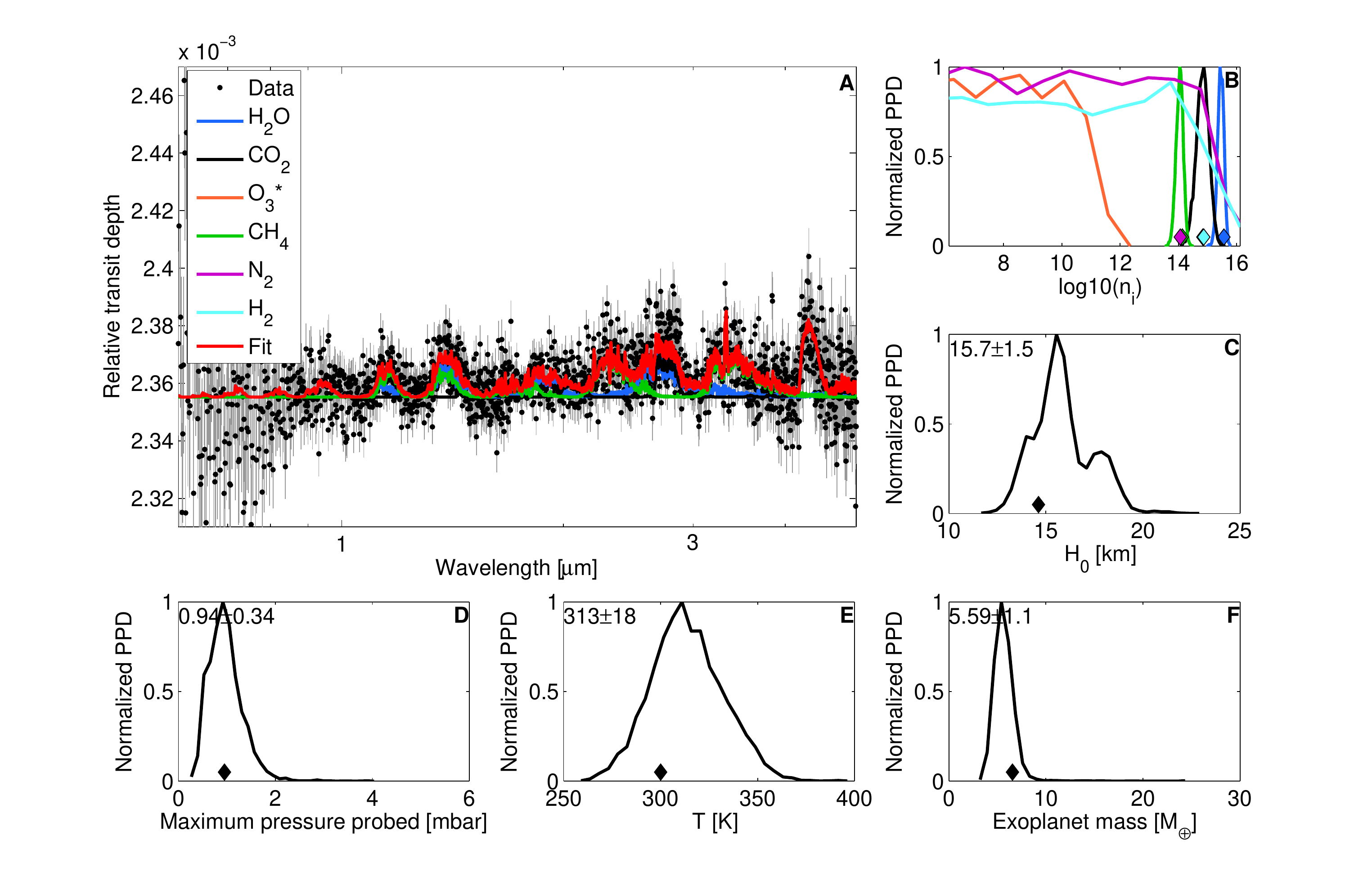}
  \end{center}
  \vspace{-0.6cm}
  \caption{\textit{MassSpec}'s application to the synthetic transmission spectrum of a super-Earth with a water-dominated atmosphere presenting a cloud deck at 1 mbar [lowest pressure level where clouds are expected \cite{Howe2012}] transiting a M1V star at 15 pc as observed with \textit{JWST} for a total of 200 hrs in-transit. The panels show the same quantities as on Fig.\,\ref{fig:MassSpec_results_in_text_ww}. Despite the presence of high and thick clouds, the atmospheric properties are retrieved with sufficient significance to yield a mass measurement with a relative uncertainty of $\sim20\%$ (i.e., twice larger than in the cloud-free scenario).}
  \label{fig:water_world_JWST_C3}
\end{figure}



\begin{figure}[!ht]
  \begin{center}
    \includegraphics[trim = 00mm 00mm 00mm 00mm,clip,width=17cm,height=!]{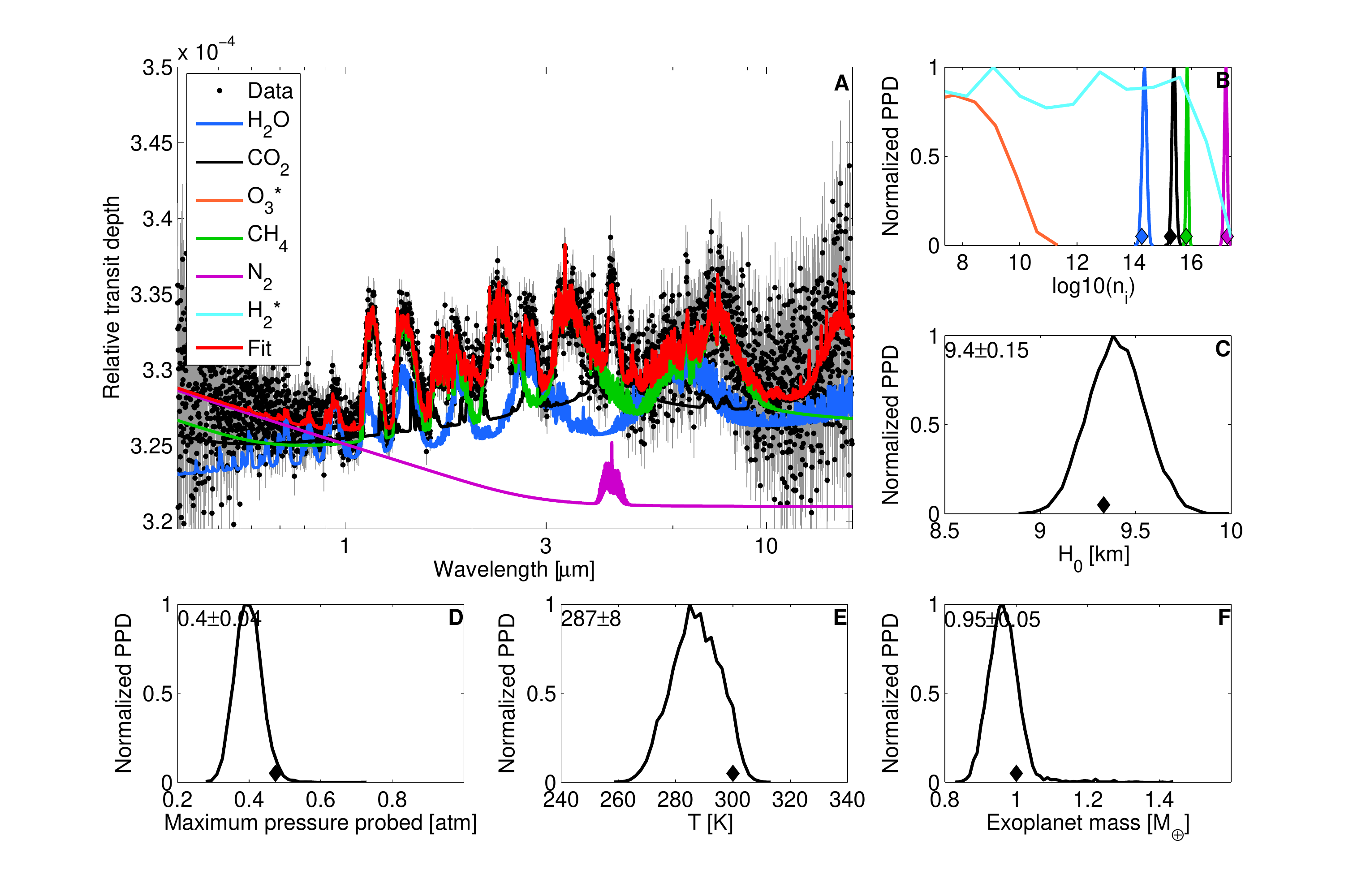}
  \end{center}
  \vspace{-0.6cm}
  \caption{\textit{MassSpec}'s application to the synthetic transmission spectrum of an Earth-sized planet with a nitrogen-dominated atmosphere transiting a M1V star at 15 pc as observed with a future-generation 20-meter space telescope for a total of 200 hrs in-transit. The panels show the same quantities as on Fig.\,\ref{fig:MassSpec_results_in_text_ww}.  
The atmospheric properties are retrieved with high significance yielding to a mass measurement with a relative uncertainty of $\sim5\%$. Note that the significant observation of the Rayleigh-scattering slope combined with the lack of $H_2-H_2$ CIA feature at 3 microns (Fig.\,\ref{fig:mini_neptune_JWST}) yields to the retrieval of nitrogen as the dominant atmospheric species.}
  \label{fig:nitrogen_world_20m}
\end{figure}


\begin{figure*}[!ht]
 \centering
  \begin{center}
    \includegraphics[trim = 00mm 00mm 00mm 00mm,clip,width=14cm,height=!]{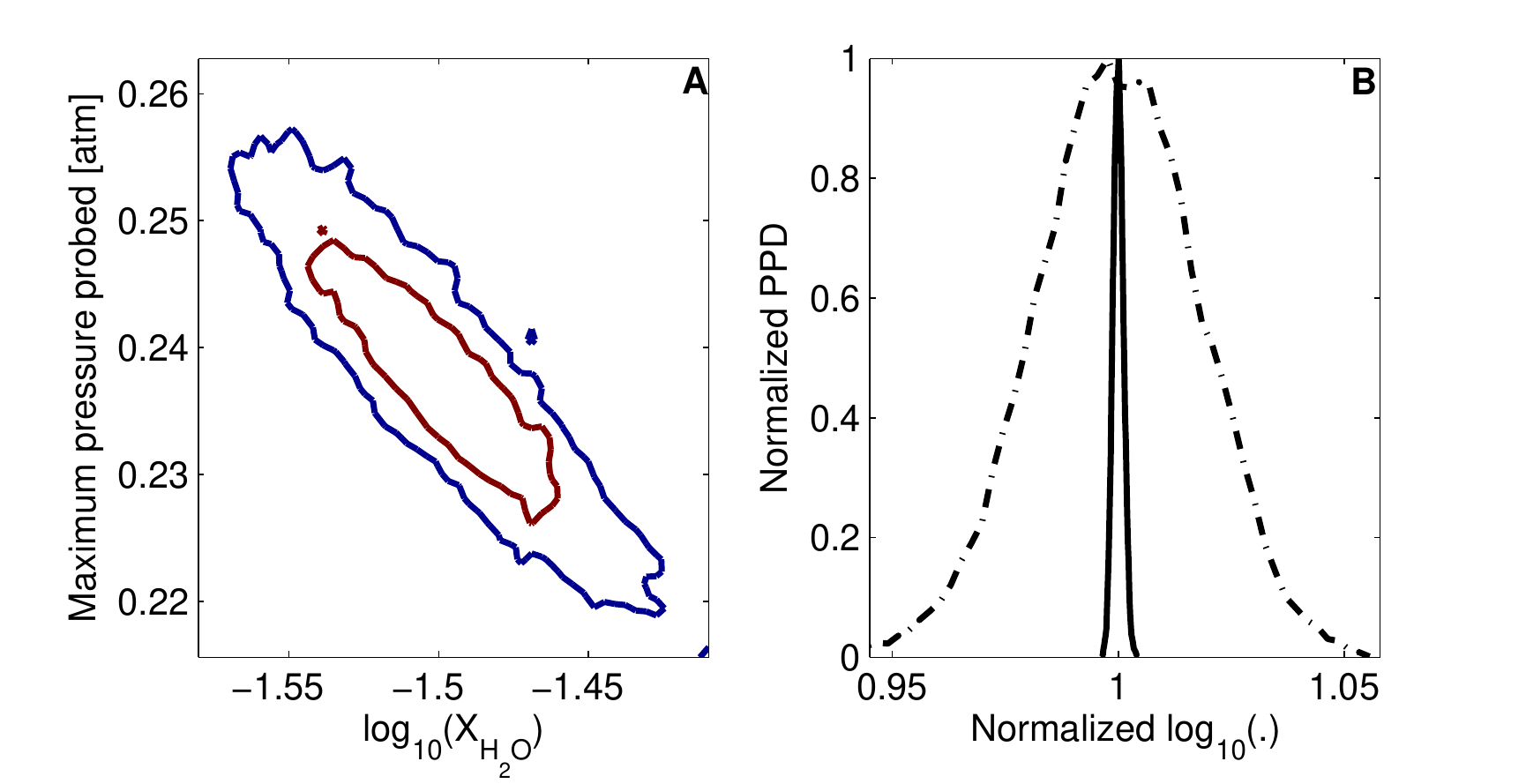}
  \end{center}
  \vspace{-0.7cm}
  \caption{The use of species' mixing ratios in atmosphere retrieval method. (\textbf{A}) Marginal posterior probability distribution (69$\%$ and 95$\%$ confidence intervals, respectively in red and blue) of maximum pressure probed and H$_2$O mixing ratio that shows the correlation between both parameters. The correlation shown between the pressure and a species mixing ratio translates the fact that the key parameter of a planet's transmission spectrum are the number densities, not the mixing ratios. Therefore, the uncertainty on mixing ratios combine the uncertainty on the pressure and the number densities. (\textbf{B}) Water's mixing ratio PPD (dot-dash line) and water's number density PPD (solid line), shifted along the x-axis for comparison. The larger extent of the water mixing ratio PPD results from the combination of the uncertainty on the number density and pressure.}
  \vspace{-0.0cm}
  \label{fig:pressuremix}
\end{figure*}

\clearpage

\begin{figure}[!p]
 \centering
  \begin{center}
    \includegraphics[trim = 00mm 00mm 00mm 00mm,clip,width=15cm,height=!]{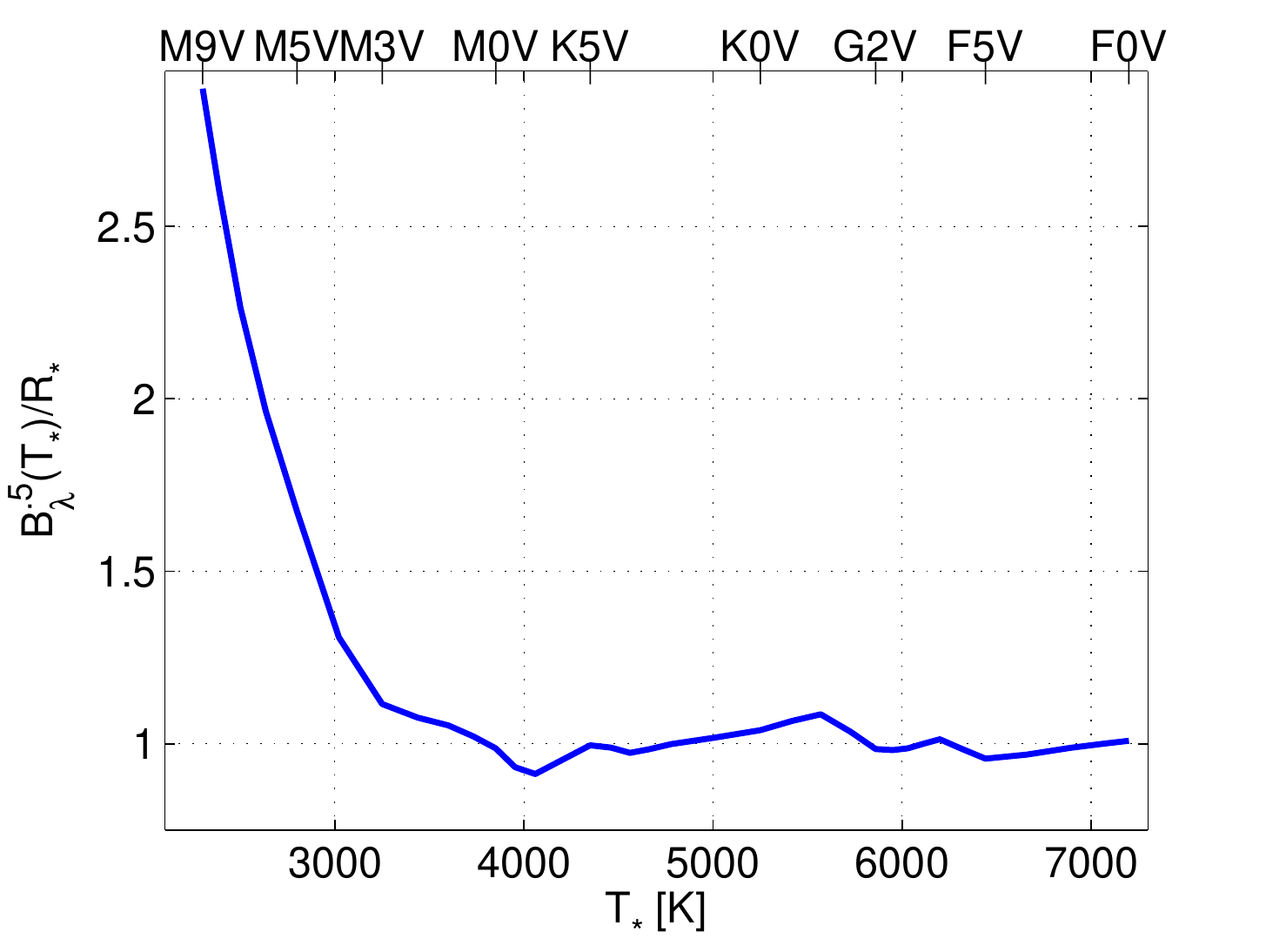}
  \end{center}
  \vspace{-0.7cm}
  \caption{The significance of in-transit planetary signals increases for planets transiting M dwarfs. The figure shows the ratio $\sqrt{B_{\lambda}(T_{\star})}/R_{\star}$---normalized for a Sun-like star---as a function of the stellar effective temperature. $\sqrt{B_{\lambda}(T_{\star})}/R_{\star}$ scales as the overall significance of an in-transit planetary signal such as a transmission spectrum for fixed planetary properties (Eq.\,\ref{eq:SNRt_general_scalinglaw_fixed_planet}). For stars with earlier spectral types than M2V, the significance is independent of the host-star type. However, the significance increase significantly towards late M dwarfs.}
  \vspace{-0.0cm}
  \label{fig:key_ratio_fixed_planet}
\end{figure}

\clearpage

\begin{figure}[!p]
 \centering
  \begin{center}
    \includegraphics[trim = 20mm 00mm 20mm 00mm,clip,width=15cm,height=!]{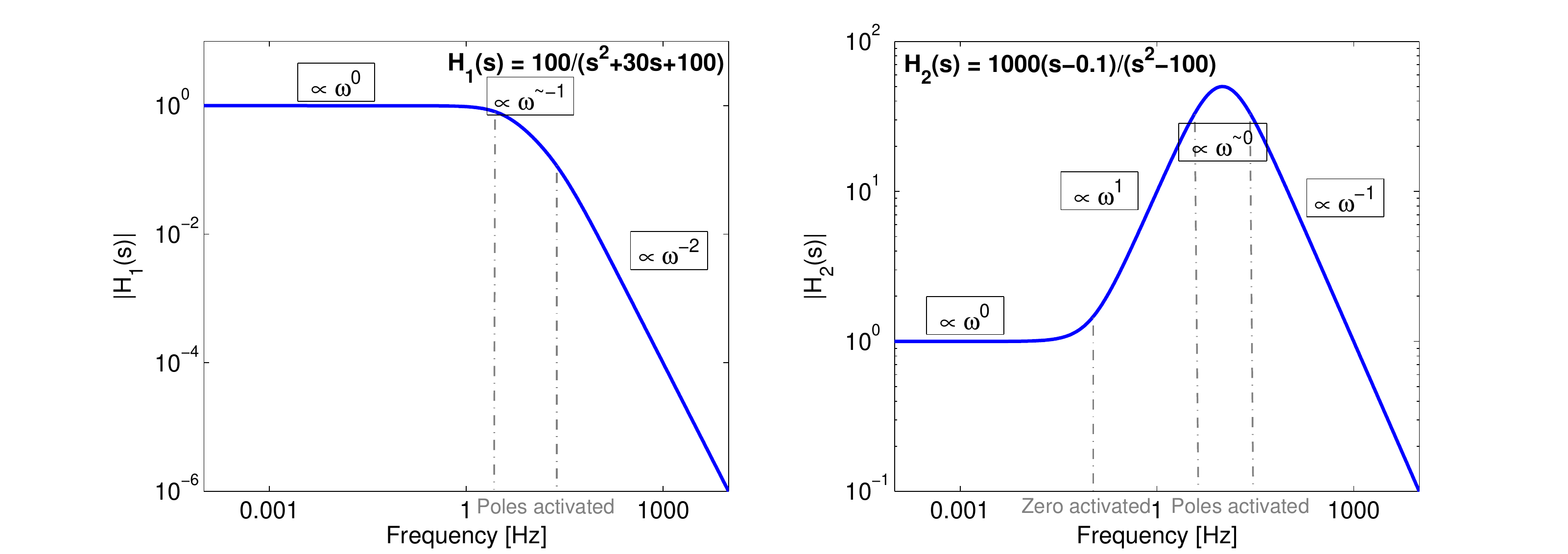}
  \end{center}
  \vspace{-0.7cm}
  \caption{Typical behavior of rational functions. Left: Transfer function of a mass-spring-dashpot system with a natural frequency of 10 Hz and a damping ratio of 1.5. The transfer function is independent of the frequency at frequency lower than the system's natural frequency. At frequency larger than the natural frequency, the two poles of the transfer function are activated and the dependency of the transfer function on the frequency changes from $\propto \omega^0$ to $\propto \omega^{-2}$ (i.e., the exponent decreases by two). Right: Transfer function with a zero at 0.1 Hz and a pair of conjugated poles at $\pm$10 Hz. The effect of the zero is to increase the exponent of the transfer function dependency on the frequency by one, while the effect of the pair of poles is to decrease it by two.}
  \vspace{-0.0cm}
  \label{fig:rational_function}
\end{figure}


\begin{thebibliography}{10}

\bibitem{Schneider2011}
J.~{Schneider}, C.~{Dedieu}, P.~{Le Sidaner}, R.~{Savalle}, I.~{Zolotukhin},
  {\it \aap\/} {\bf 532}, A79 (2011).

\bibitem{Batalha2013}
N.~M. {Batalha}, {\it et~al.\/}, {\it \apjs\/} {\bf 204}, 24 (2013).

\bibitem{Stamenkovic2012}
V.~{Stamenkovi{\'c}}, L.~{Noack}, D.~{Breuer}, T.~{Spohn}, {\it \apj\/} {\bf
  748}, 41 (2012).

\bibitem{Collier2010}
A.~{Collier Cameron}, {\it et~al.\/}, {\it \mnras\/} {\bf 407}, 507 (2010).

\bibitem{Mislis2012}
D.~{Mislis}, R.~{Heller}, J.~H.~M.~M. {Schmitt}, S.~{Hodgkin}, {\it \aap\/}
  {\bf 538}, A4 (2012).

\bibitem{Fabrycky2010}
D.~C. {Fabrycky}, {\it {Non-Keplerian Dynamics of Exoplanets}\/} (University of
  Arizona Press, 2010), pp. 217--238.

\bibitem{Faigler2011}
S.~{Faigler}, T.~{Mazeh}, {\it \mnras\/} {\bf 415}, 3921 (2011).

\bibitem{Agol2005}
E.~{Agol}, J.~{Steffen}, R.~{Sari}, W.~{Clarkson}, {\it \mnras\/} {\bf 359},
  567 (2005).

\bibitem{Holman2005}
M.~J. {Holman}, N.~W. {Murray}, {\it Science\/} {\bf 307}, 1288 (2005).

\bibitem{Steffen2013}
J.~H. {Steffen}, {\it et~al.\/}, {\it \mnras\/} {\bf 428}, 1077 (2013).

\bibitem{Barstow2013}
J.~K. {Barstow}, {\it et~al.\/}, {\it \mnras\/} {\bf 430}, 1188 (2013).

\bibitem{science1note5}
We also show that Eq.\,\ref{eq:h_eff_in_text} can be rewritten as
  $\tau(h_{eff}(\lambda),\lambda) \triangleq \tau_{eq} = e^{-\gamma_{EM}}$
  meaning that the slant-path optical depth at the apparent height is a
  constant (Fig.\,\ref{fig:transmission_spectrum_basics}, panel C)---this
  extends previous numerical observations that $\tau_{eq}\approx0.56$ in some
  case \cite{LecavelierDesEtangs2008}. Therefore $\tau_{eq} = \lim_{n \to
  +\infty} n\prod_{k = 1}^n e^{-1/k}$ ($\approx 0.56146$).

\bibitem{Seager2010}
S.~{Seager}, {\it {Exoplanet Atmospheres: Physical Processes}\/} (Princeton
  University Press, 2010).

\bibitem{Euler1740}
L.~{Euler}, {\it Comm. Petropol.\/} {\bf 7}, 150 (1740).

\bibitem{LecavelierDesEtangs2008}
A.~{Lecavelier Des Etangs}, F.~{Pont}, A.~{Vidal-Madjar}, D.~{Sing}, {\it
  \aap\/} {\bf 481}, L83 (2008).

\bibitem{Madhusudhan2009}
N.~{Madhusudhan}, S.~{Seager}, {\it \apj\/} {\bf 707}, 24 (2009).

\bibitem{Pont2008}
F.~{Pont}, H.~{Knutson}, R.~L. {Gilliland}, C.~{Moutou}, D.~{Charbonneau}, {\it
  \mnras\/} {\bf 385}, 109 (2008).

\bibitem{Wright2011}
J.~T. {Wright}, {\it et~al.\/}, {\it \pasp\/} {\bf 123}, 412 (2011).

\bibitem{Kaltenegger2009}
L.~{Kaltenegger}, W.~A. {Traub}, {\it \apj\/} {\bf 698}, 519 (2009).

\bibitem{Demory2013}
B.-O. {Demory}, {\it et~al.\/}, {\it \apjl\/} {\bf 776}, L25 (2013).

\bibitem{Barstow2013b}
J.~K. {Barstow}, S.~{Aigrain}, P.~G.~J. {Irwin}, L.~N. {Fletcher}, J.-M. {Lee},
  {\it \mnras\/} {\bf 434}, 2616 (2013).

\bibitem{Howe2012}
A.~R. {Howe}, A.~S. {Burrows}, {\it \apj\/} {\bf 756}, 176 (2012).

\bibitem{Seager2007}
S.~{Seager}, M.~{Kuchner}, C.~A. {Hier-Majumder}, B.~{Militzer}, {\it \apj\/}
  {\bf 669}, 1279 (2007).

\bibitem{Fortney2007}
J.~J. {Fortney}, M.~S. {Marley}, J.~W. {Barnes}, {\it \apj\/} {\bf 659}, 1661
  (2007).

\bibitem{Rogers2011}
L.~A. {Rogers}, P.~{Bodenheimer}, J.~J. {Lissauer}, S.~{Seager}, {\it \apj\/}
  {\bf 738}, 59 (2011).

\bibitem{Fortney2005}
J.~J. {Fortney}, {\it \mnras\/} {\bf 364}, 649 (2005).

\bibitem{Borysow2002}
A.~{Borysow}, {\it \aap\/} {\bf 390}, 779 (2002).

\bibitem{Rothman2009}
L.~S. {Rothman}, {\it et~al.\/}, {\it \jqsrt\/} {\bf 110}, 533 (2009).

\bibitem{Liu2001}
Y.~{Liu}, J.~{Lin}, G.~{Huang}, Y.~{Guo}, C.~{Duan}, {\it Journal of the
  Optical Society of America B Optical Physics\/} {\bf 18}, 666 (2001).

\bibitem{Sneep2005}
M.~{Sneep}, W.~{Ubachs}, {\it \jqsrt\/} {\bf 92}, 293 (2005).

\bibitem{Benneke2012}
B.~{Benneke}, S.~{Seager}, {\it \apj\/} {\bf 753}, 100 (2012).

\bibitem{Hu2013}
R.~{Hu}, S.~{Seager}, W.~{Bains}, {\it \apj\/} {\bf 769}, 6 (2013).

\bibitem{Boker2010}
T.~{B{\"o}ker}, J.~{Tumlinson}, {NIRSpec Operations Concept Document - James
  Webb Space Telescope}, {\it Tech. Rep. ESA-JWST-TN-0297 (JWST-OPS-003212)\/},
  ESA (2010).

\bibitem{Deming2009}
D.~{Deming}, {\it et~al.\/}, {\it \pasp\/} {\bf 121}, 952 (2009).

\bibitem{Charbonneau2009}
D.~{Charbonneau}, {\it et~al.\/}, {\it \nat\/} {\bf 462}, 891 (2009).

\bibitem{Zsom2013}
A.~{Zsom}, S.~{Seager}, J.~{de Wit}, V.~{Stamenkovic}, {\it \apj\/} {\bf 778},
  109 (2013).

\bibitem{Miller-Ricci2009}
E.~{Miller-Ricci}, S.~{Seager}, D.~{Sasselov}, {\it \apj\/} {\bf 690}, 1056
  (2009).

\bibitem{Murray2010}
C.~D. {Murray}, A.~C.~M. {Correia}, {\it {Keplerian Orbits and Dynamics of
  Exoplanets}\/} (University of Arizona Press, 2010), pp. 15--23.

\bibitem{Cox2000}
A.~N. {Cox}, C.~A. {Pilachowski}, {\it Physics Today\/} {\bf 53}, 100000
  (2000).

\end{thebibliography}
\end{document}